\documentclass[twocolumn]{aa}

\usepackage{amsmath,natbib, appendix}
\usepackage{longtable}
\usepackage{graphicx}
\usepackage{txfonts}

\begin{document}

\title{The first molecules in the intergalactic medium and halos of the Dark Ages and Cosmic Dawn}
\titlerunning{The First Molecules}

\author{B. Novosyadlyj\inst{1,2}, Yu. Kulinich\inst{1}, B. Melekh\inst{1}, V. Shulga\inst{2,3}}
\authorrunning{Novosyadlyj et al.} 
\institute{$^1$Ivan Franko National University of Lviv, Kyryla i Methodia str., 8, Lviv, 79005, Ukraine, \\
$^2$Jilin University, Qianjin Street 2699, Changchun, 130012, P.R.China,\\
$^3$Institute of Radio Astronomy of NASU, 4 Mystetstv str., 61002 Kharkiv, Ukraine}

\abstract
{We study the formation and destruction of the first molecules at the epochs of the Dark Ages and Cosmic Dawn to evaluate the luminosity of the protogalaxy clumps  (halos) in the molecular lines.  The cosmological recombination is described using the  RecFast model of an effective three-level atom, while the chemistry of the molecules is examined using the relevant basic  kinetic equations. We then studied the effect of collisional and radiative excitation of molecules on the intensity of molecular emission  in both warm and hot halos. Using the Planck data on the reionization of the intergalactic medium at $z\sim6-8,$ we evaluated the upper limits of the light energy density for four models of thermal light from the first sources that appeared in the Cosmic Dawn epoch. Assuming that in the halos, the light energy density may  essentially be even higher, we estimated the impact of the light from the first sources (the first light) on the formation and destruction of the first molecules in them as well as between them. We show that the molecules H$_2$ and HD are destroyed by photodissociation processes shortly before the full reionization in the inter-halo medium, in the medium of both types of halos and for all models of the first light. At the same time, the number density of  helium hydride ions, HeH$^+$, shows essentially more complicated dependences on the kinetic temperature of halos and the models of the first light. These features characterizing the abundance of molecules also determine the intensity of the halos luminescence during their evolution. Furthermore, we calculated the evolution of the brightness temperature of the individual halo in the rotational lines of H$_2$, HD and HeH$^+$ molecules relative to the temperature of the cosmic microwave background at redshifts corresponding to the Dark Ages and Cosmic Dawn epochs. It does not exceed the microkelvin, but its detection may be an important source of information about the physical processes taking place at the beginning of the formation of the first stars and galaxies at the epochs of the Dark Ages and Cosmic Dawn.}


\maketitle

\section{Introduction}
\small{
\begin{table*}[!ht]
\caption{Main chemical reactions and references for their rates.}
\label{tab1}
\centering
\begin{tabular}{l l l|l l l}
\hline\hline
&reaction&references&&reaction&references\\
\hline
(H1)&$\mathrm{H^+ + e^- \rightarrow H + \gamma}$&Appendix A&(H2)&$\mathrm{H + \gamma\rightarrow H^+ + e^-}$&Appendix A\\
(H3)&$\mathrm{H + e^- \rightarrow H^- + \gamma}$&\cite{deJong1972}&(H4)&$\mathrm{H^- + \gamma \rightarrow H + e^-}$&\cite{deJong1972}\\
(H5)&$\mathrm{H^- + H \rightarrow H_2 + e^-}$&\cite{Launey1991}&(H6)&$\mathrm{H^- + H^+ \rightarrow H_2^+ +e^-}$&\cite{Poulaert1978} \\
(H7)&$\mathrm{H^- + H^+ \rightarrow H +H}$&\cite{Moseley1970} &(H8)&$\mathrm{H + H^+ \rightarrow H_2^+ + \gamma}$&\cite{Stancil1993}\\
(H9)&$\mathrm{H_2^+ + \gamma \rightarrow H + H^+}$&\cite{Stancil1994}&(H10)&$\mathrm{H_2^+ + H \rightarrow H_2 + H^+}$&\cite{Karpas1979} \\
(H11)&$\mathrm{H_2^+ + e^- \rightarrow H + H}$&\cite{Schneider1994}&(H12)&$\mathrm{H_2^+ + \gamma \rightarrow 2H^+ + e^-}$&\cite{Bates1968}\\
(H15)&$\mathrm{H_2 + H^+ \rightarrow H_2^+ + H}$&\cite{Holliday1971}&(H18)&$\mathrm{H_2 + \gamma \rightarrow H_2^+ + e^-}$&\cite{ONeil1978}\\
(H23)&$\mathrm{H_2 + \gamma \rightarrow H + H}$&Appendix B&&&\\
\hline
(D1)&$\mathrm{D^+ + e^- \rightarrow D + \gamma}$&Appendix A&(D2)&$\mathrm{D + \gamma \rightarrow D^+ + e^-}$&Appendix A\\
(D3)&$\mathrm{D + H^+ \rightarrow D^+ + H}$&\cite{Watson1978}&(D4)&$\mathrm{D^+ + H \rightarrow D+H^+}$&\cite{Watson1978}\\
(D8)&$\mathrm{D^+ + H_2 \rightarrow H^+ + HD}$&\cite{Smith1982}&(D10)&$\mathrm{HD + H^+ \rightarrow H_2 + D^+}$&\cite{Smith1982}\\
(D25)&$\mathrm{HD + \gamma \rightarrow D + H}$&Appendix B&&&\\
\hline
(He1)&$\mathrm{He^{++} + e^- \rightarrow He^+ + \gamma}$&\cite{Verner1996}&(He2)&$\mathrm{He^+ + \gamma\rightarrow He^{++} + e^-}$& \cite{Osterbrock1989}\\
(He3)&$\mathrm{He^{+} + e^- \rightarrow He + \gamma}$& \cite{Verner1996}&(He4)&$\mathrm{He + \gamma\rightarrow He^{+} + e^-}$&\cite{Osterbrock1989}\\
(He8)&$\mathrm{He + H^+ \rightarrow HeH^+ + \gamma}$&\cite{Roberge1982}&(He10)&$\mathrm{He^+ + H \rightarrow HeH^+ + \gamma}$&\cite{Zygelman1990}\\
(He11)&$\mathrm{HeH^+ + H \rightarrow He + H_2^+}$&\cite{Karpas1979}&(He12)&$\mathrm{HeH^+ + e^- \rightarrow He + H}$&\cite{Yousif1989}\\
(He14)&$\mathrm{HeH^+ + \gamma \rightarrow He + H^+}$&\cite{Roberge1982}&(He15)&$\mathrm{HeH^+ + \gamma \rightarrow He^+ + H}$&\cite{Roberge1982}\\
\hline
\end{tabular}
\end{table*}}

%%%%%%%%%%%%%%%%%%%%%%
The upcoming decade promises to be a time of significant breakthroughs in the observational investigation exploring the epochs of the Dark Ages and Cosmic Dawn. These studies are expected to fill the gap in our understanding of how the first stars and galaxies formed. A few teams of cosmologists and observers are competing to detect the signal in the hyperfine 21 cm line of atomic hydrogen that is redshifted according to these epochs. The first detection of the absorption in this line at $z=15-20$ was announced by the EDGES\footnote{https://www.haystack.mit.edu/ast/arrays/Edges/} team a few years ago \citep{Bowman2018,Hills2018}. Other sources of important information about the formation of the first stars and galaxies are the first molecules, which absorb the cosmic microwave background radiation (CMBR) in their ro-vibrational transitions in cold baryonic matter clouds, while emitting additional radiation in hot ones or resonantly scattering it in moving clouds \citep{Dubrovich1977,Maoli1994,Maoli1996,Kamaya2002,Ripamonti2002,Kamaya2003,Omukai2003,Mizusawa2004,Mizusawa2005,Dubrovich2008,Liu2019,Novosyadlyj2020,Maio2022}. Unfortunately, predictions of the amplitudes of such distortions of CMBR are at the level of microkelvins and too low for their detection by current telescopes. The upper observational limits obtained over the last few decades for such signals are at the level of a few microkelvins \citep{deBernardis1993,Gosachinskij2002,Persson2010}. The detectability of primordial pre-galactic clouds in the molecular lines has been discussed by some authors over the two past decades \citep{Kamaya2002,Ripamonti2002,Kamaya2003,Omukai2003,Mizusawa2004,Mizusawa2005,Liu2019,Novosyadlyj2020,Maio2022} and a marginal possibility has been predicted for modern and projected telescopes such as  ALMA\footnote{https://www.almaobservatory.org/en/home/}, JWST\footnote{https://www.jwst.nasa.gov/}, SAFIR Observary\footnote{https://safir.jpl.nasa.gov/}, SPICA\footnote{https://www.ir.isas.jaxa.jp/SPICA/}, and OST\footnote{https://asd.gsfc.nasa.gov/firs/}.

In ours previous papers \citep{Novosyadlyj2017,Novosyadlyj2018,Novosyadlyj2020,Kulinich2020} we studied the formation and destruction of the first molecules, H$_2$, HD, and HeH$^+$, in the halos of the Dark Ages and their luminescence in the rotational lines of these molecules. Here, we elucidate the matter of how the radiation of the first sources,  arising at Cosmic Dawn, influences the abundances of the first molecules and their luminescence in the same rotational lines. For simplicity, we suppose the thermal energy distribution of the first sources and consider the different evolution of its spectral density and radiation temperature at the moment of cosmological reionization, $z_{rei}$. The latter we assume to be between $6\le z\le8,$ according to the observational data coming from the measurements of the polarization of CMBR at low spherical harmonics \citep{Planck2020a,Planck2020}, Gunn-Peterson trough in the spectra of high-redshift quasars \citep{Bouwens2015,Banados2018,Davies2018}, and the Lyman Break galaxies \citep{Mason2018}.
 
All computations in the paper were performed for consistent values of the main parameters of the cosmological model following the final data release of the Planck Space Observatory \citep{Planck2020}: the Hubble constant $H_0=67.4$ km/s/Mpc, the mean density of baryonic matter in the units of critical one $\Omega_b=0.0493$, the mean density of dark matter $\Omega_{dm}=0.2657$, the mean density of dark energy $\Omega_{de}=0.685$, its equation of state parameter $w_{de}=-1.03$, the spectral index of the scalar mode of cosmological perturbations $n_s=0.965$ and the matter fluctuation amplitude $\sigma_8=0.811$, and the current temperature of cosmic microwave radiation $T_0=2.7255$ K. We use also the primordial helium abundance $Y_p=0.2446$ \citep{Peimbert2016} and deuterium fraction $y_{Dp}=2.527\cdot10^{-5}$ \citep{Cooke2018}, which  are in good agreement with the posterior means from \cite{Planck2020}.

The outline of this paper is as follows. In Section 2, we present the network of chemical reactions leading to the formation and destruction of molecules and the recombination and photoionization of atoms as well as the corresponding kinetics equations and energy balance. In Section 3, we describe the models of the energy distribution for the first sources of light (the first light) and estimate the value of the dilution coefficient for each from them. In Section 4, we compute the kinetics of chemical reactions in the inter-halo medium and analyze the influence of the first light on the number density of the first molecules H$_2$, HD and HeH$^+$. The intra-halo chemistry and luminosity of halos in the ro-vibrational molecular lines are analyzed in Section 5. To do this, together with the equations of the kinetics of chemical reactions and the kinetics of excitations and de-excitations of the ro-vibrational energy levels of the first molecules, we solve the equations for the evolution of the density and temperature of matter. The top-hat halos with hot and warm baryonic gas inside and with masses that are typical for $\Lambda$CDM/$\Lambda$WDM models of the Universe are considered. As a result, we obtain the evolution of the number density of molecules H$_2$, HD, and HeH$^+$ at different stages of halo formation up to the reionization epoch for different models of the first light. Our conclusions are given in Section 6. Useful approximations for the rates of photorecombination and photoionization for HI and DI are presented in Appendix A, while the rates of photodissociation for molecules H$_2$ and HD are presented in Appendix B.
 
\section{Chemical network}

We analyze the formation and destruction of the most abundant molecules in the inter-proto-galaxy medium of the Dark Ages and Cosmic Dawn epochs: H$_2$, HD, and HeH$^+$. The minimal set of reactions important for epochs starting from cosmological recombination, Dark Ages, Cosmic Dawn, and reionization is presented in Table \ref{tab1}. The rates and fitting formulas for them are taken from the sources cited in the last column. Most of fitting formulas are collected in Tables 1-3 of \cite{Abel1997}, Tables 1-3 of \cite{Galli1998}, and Table B.1 of \cite{Schleicher2008}. We also keep here the notations of reactions (in parentheses) introduced there. In comparison with a minimal set of reactions important for the Dark Ages epoch \citep{Galli1998,Novosyadlyj2018}, we have added the important reactions for the Cosmic Dawn epoch, when the first light would ionize atoms and molecules and then photodissociate the molecules. These are: H6, H11, H12, H18, H23, D25, He10, He12, and He15. Since the ionization of hydrogen is important for the formation and destruction of molecules, we propose the more accurate analytical approximation of effective cross-sections of hydrogen photoionization from the base level (case A) and the first metastable 2s one (case B) in Appendix A. The direct reaction of photodissociation H23, as well as that of D25, becomes crucial for the destruction of hydrogen and hydrogen deuteride molecules by first light before reionization \citep{Glover2007,Schleicher2008}. We propose the new approximation of the rate of photodissociation of H$_2$ (HD) using recent and accurate computations of effective cross-section for this purpose from \cite{Heays2017}, presented in Appendix B.

We computed the evolution of fractions, $x_{\rm i}$, of neutral atoms, molecules, and ions from the cosmological recombination up to hydrogen reionization at $z_{rei}$. The index ``${\rm i}$'' notes any atom, molecule, or ion presented in Table \ref{tab1}. The kinetic equations for chemical reactions in the general form are as follows \citep{Puy1993,Galli1998,Vonlanthen2009,Novosyadlyj2018}:
\begin{eqnarray}
-(1+z)H\frac{dx_i}{dz}&=&\sum_{mn}C_{mn}^{(i)}f_{m} f_{n} x_{m}x_{n}+\sum_{m}R_{m\gamma}^{(i)}f_{m} x_{m}\nonumber\\
&-&\sum_{j}C_{ij}f_{i} f_{j} x_{i}x_{j}-\alpha_{i}f_{i}x_{i},
\label{kes}
\end{eqnarray}
where $z$ is redshift; $C_{mn}^{(i)}$ is the rate of collision reaction with reactants, ${m}$ and ${n,}$ which leads to the formation of the atom, molecule, ion, ${i}$; $R_{m\gamma}^{(i)}$ is the rate of radiative reaction for reactant, ${m,}$ which leads to the formation of an atom, molecule, ion, ${i}$; $C_{in}$ is the rate of  destruction of ${i}$-reactant by collision with an n-reactant; and $\alpha_i$ is the photorecombination rate of the reactant, $i$. The total fraction of the main chemical elements in the unit number density of hydrogen, denoted as $f_{m}$, namely, $f_{He}=n_{He}/n_{H}$ for a reactant, ${m,}$ containing helium; $f_{D}=n_{D}/n_{H}$ for a reactant, ${m,}$ containing deuterium; and $f_{H}\equiv1$ for reactant, ${m,}$ containing hydrogen only. For chemical species containing only hydrogen, the fraction, ${m,}$ is $x_{m}=n_{m}/n_{H}$, where $n_{m}$ is the number density of species, ${m}$, and $n_{H}$ is the total number density of hydrogen; for species containing deuterium and helium,  $x_{m}=n_{m}/n_{D}$ and $x_{m}=n_{m}/n_{He}$, respectively, where $n_{D}$ and $n_{He}$ are the total number densities of deuterium and helium.
 
Equations (\ref{kes}) for the reactions presented in Table \ref{tab1} together with equations of cosmological recombinations for hydrogen, deuterium, and helium \citep{seager1999,seager2000,Novosyadlyj2017} compose a system of kinetic equations which we solve numerically together with the equations of expansion of the Universe and energy balance for the baryonic component:
\begin{eqnarray}
&&H=\sqrt{\Omega_r(1+z)^4+\Omega_m(1+z)^3+\Omega_{de}(1+z)^{3(1+\omega_{de})}},  \label{H}\\
&&\frac{dT_{b}}{dz}=\frac{2T_{b}}{1+z}+\frac{8\sigma_Ta_rT_r^4}{3m_ecH(1+z)}\frac{x_e}{1+f_D+f_{He}+x_e}\left(T_b-T_r\right),\nonumber \\
&&\hskip1.3cm+\alpha_{fl}\frac{8\sigma_Ta_rT_{fl}^4}{3m_ecH(1+z)}\frac{x_e}{1+f_D+f_{He}+x_e}\left(T_b-T_{fl}\right).\label{Tb}
\end{eqnarray}
where $\sigma_T$ is the Thomson scattering cross-section, $a_r$ is the radiative constant, and $m_e$ is the mass of an electron. We set the initial conditions for these reactions at the early epoch before the cosmological recombination when all ingredients were completely ionized and a Saha approximation was applicable (see details in Appendix A in \cite{Novosyadlyj2017}).  
The publicly available codes RecFast\footnote{http://www.astro.ubc.ca/people/scott/recfast.html} and  DDRIV1\footnote{http://www.netlib.org/slatec/src/ddriv1.f} have been used in the general code CDhalo.f, which was designed for integrating the system of equations (\ref{kes}) in the expanding Universe over cosmological recombination, Dark Ages, and Cosmic Dawn, when the first light becomes important for the ionization and dissociation of atoms and molecules. The last equation (\ref{Tb}) is used at $z\le850$, at higher $z$  $T_b=T_r=T_0(1+z)$. It takes into account the main mechanisms of cooling or heating of diluted gas in the Dark Ages and Cosmic Dawn: adiabatic cooling and heating by CMBR at the early epoch and by the first light in the late ones.
 
\section{Models of the first light}
In the epoch of Cosmic Dawn, the radiation of the first sources (warm or hot halos, the first dwarf galaxies with Pop III stars, and shock waves), referred to here as the first light, starts to impact the kinetics of the chemical reactions, exciting the ro-vibrational levels of molecules; and at the end of Cosmic Dawn, it starts to ionize the hydrogen, marking the beginning of the epoch of reionization. To indicate the moment of reionization, we assume that this happens when the fraction of ionized hydrogen reaches a value of 0.5. According to the Planck 2018 results on the CMBR anisotropy, this takes place at $6\le z\le8$ (see Fig. 45 in \cite{Planck2020}). Other observations, such as the Gunn-Peterson trough in the spectra of high-redshift quasars \citep{Bouwens2015,Banados2018,Davies2018} and Lyman Break galaxies \citep{Mason2018}, support this value. Thus, here we analyze   models of the first light that satisfy the following condition: $x_{HI}=x_{HII}=0.5$ at $6\le z\le8$.
\begin{figure}[ht!]
\includegraphics[width=0.5\textwidth]{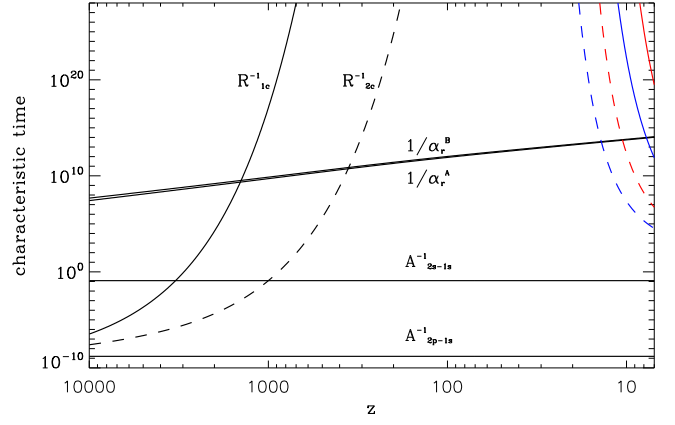}
\caption{Characteristic lifetime of the hydrogen atom in the excited states 2s and 2p ($A^{-1}_{2p-1s}$, $A^{-1}_{2s-1s}$), time intervals between following recombinations to the base and excited levels ($1/\alpha^A_r$, $1/\alpha^B_r$), and the resulting photoionization ($R^{-1}_{1c}$, $R^{-1}_{2c}$) by CMBR (at $z>200$) as well as the first light photons ($z<30$). The computations are made for the cosmological model with Planck 2018 parameters and thermal models of the first light.}
\label{tHCD}
\end{figure}

To analyze the allowable levels of illumination in the inter-proto-galaxy medium of the Cosmic Dawn epoch, we assume that sources of the first light are thermal. We consider here the thermal models of the first light described by the Planck function with temperature, $T_{fl}$, which is some smooth function of redshift and dilution coefficient $\alpha_{fl}$: $I_{fl}(\nu)=\alpha_{fl}B_{\nu}(T_{fl})$.
\begin{figure}[ht!]
\includegraphics[width=0.49\textwidth]{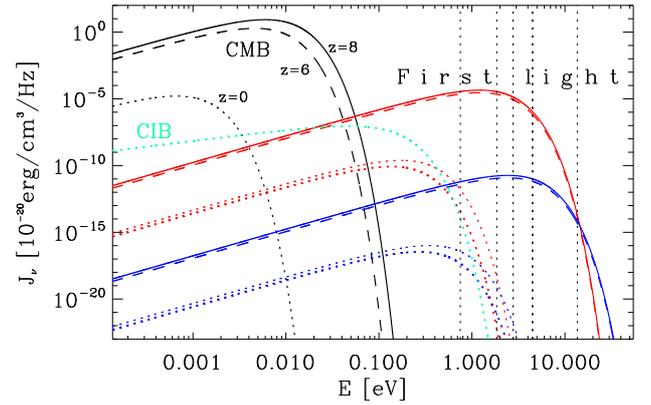}
\caption{Spectral energy density of the first light and CMBR at $z=8$ (solid lines) and at $z=6$ (dashed lines) when the ionization of hydrogen is $\approx0.5$ for models Ia and Ib. The vertical dotted lines show the thresholds of dissociation of molecules H$^-$, HeH$^+$, H$_2^+$, H$_2$, and HD (overlapped with H$_2$) and the ionization of HI, correspondingly. The dotted lines show these spectral energy densities at the current epoch. The spectral energy density of cosmic infrared background (dotted green line) is shown for comparison. In the models Ic and Id, the corresponding lines are lower by the ratios of dilution coefficients.}
\label{flI}
\end{figure}

So, in each moment of the Dark Ages and Cosmic Dawn, the total spectral energy density of radiation is the sum of the CMBR energy density and the first light one:
\begin{equation}
 J_{\nu}=\frac{4\pi}{c}\left[B_{\nu}(T_{CMB})+\alpha_{fl}B_{\nu}(T_{fl})\right].
\end{equation}
Since the rates of ionization or dissociation of atoms and molecules are computed by integration over the energy of product of cross-section and intensity per photon energy (see eq. (A1)) in Appendix), they can be presented as the following sum:
\begin{equation}
R_{i\gamma}=R_{i\gamma}(T_{CMB})+\alpha_{fl}R_{i\gamma}(T_{fl}). 
\end{equation}

\begin{figure*}[ht!]
\includegraphics[width=0.49\textwidth]{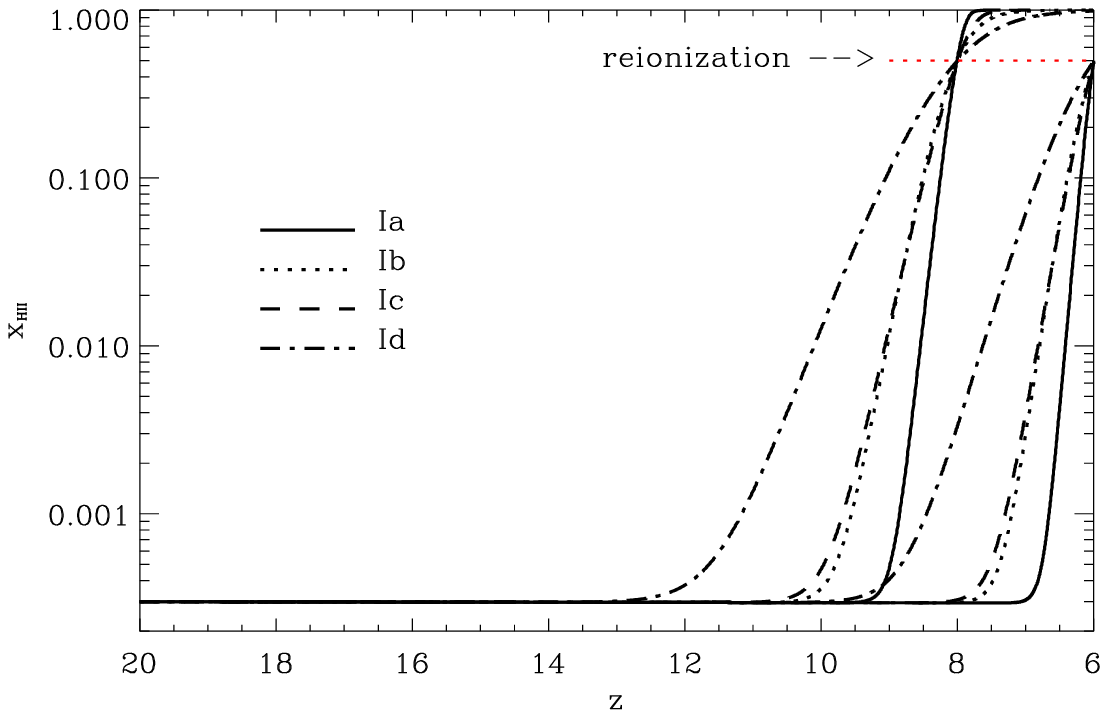}
\includegraphics[width=0.49\textwidth]{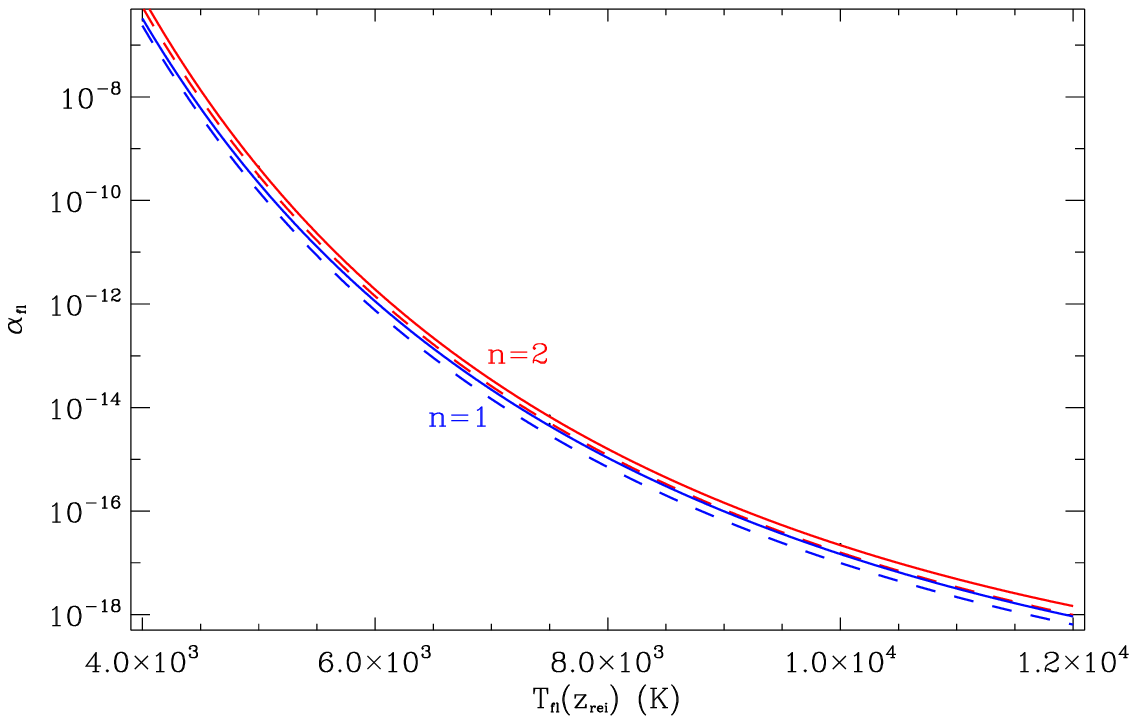}
\caption{Reionization of hydrogen by the first light at the end of the Cosmic Dawn epoch. Left: Evolution of the ionized hydrogen fraction $x_{\rm HII}$ for the first light models Ia-Id, which provide $x_{\rm HII}\approx0.5$ at $6\le z\le8$. Right: Ranges of dilution coefficient against the temperature of thermal radiation for two evolution models of the first light.}
\label{xHII}
\end{figure*}

First, we estimated which of the cases of photoionization-recombination, A or B, is realized in the Dark Ages and Cosmic Dawn. For this purpose, we compared the characteristic lifetime of the hydrogen atom in the excited states, 2s and 2p ($A^{-1}_{2p-1s}$, $A^{-1}_{2s-1s}$), with the time intervals between following photo recombinations to the ground state and excited levels ($1/\alpha^A_r$, $1/\alpha^B_r$) and photoionization from them by CMBR, and the first light photons ($R^{-1}_{1c}$, $R^{-1}_{2c}$). The collision excitation, ionization, and recombination of atomic hydrogen are insignificant in these epochs \citep{Peebles1968,Matsuda1971,Grachev1991,Abel1997} and we omitted them in this estimation. Since the $Ly_\alpha$ line is narrow, the medium is rarefied and expanding, while the $Ly_\alpha$ photons escape from the processes of scattering due to the Doppler shift of the frequency (ibid). Thus, the comparison of characteristic times is reasonable. The results are shown in Fig.~1. The characteristic times of photoionization by CMBR are shown as black  lines at $z>200$, and those by the first light are shown by color lines at $z<30$. 

\begin{figure}[ht!]
\includegraphics[width=0.49\textwidth]{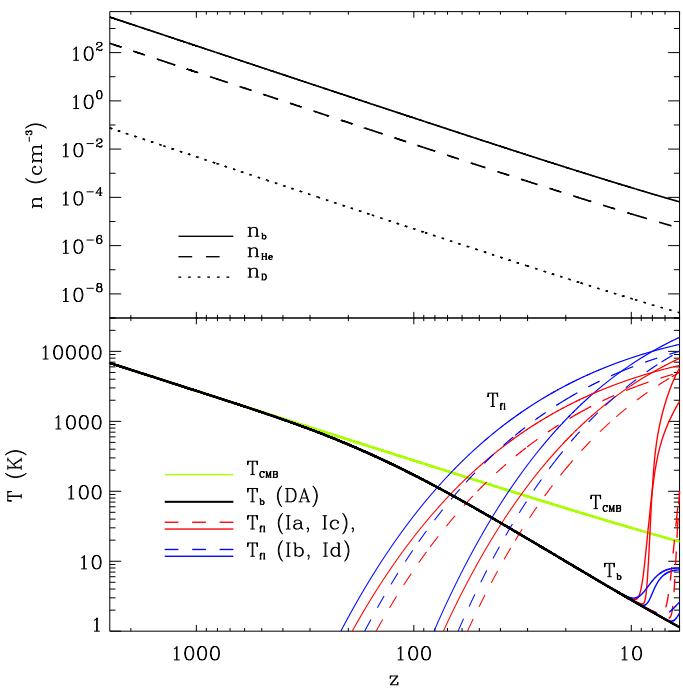} 
\caption{Evolution of the number density of H, D and He (top) and temperature of CMBR and baryonic matter through the Dark Ages to Cosmic Dawn (bottom).}
\label{roT}
\end{figure}

Case B is realized when the characteristic time of photoionization is lower or close to the lifetime of hydrogen atoms on the metastable level of 2s. We can see that this condition is satisfied at $z\gtrsim1000$ when the temperature of the CMBR was $\gtrsim3000$ K. The energy density of the first light that appeared at Cosmic Dawn, increases but does not achieve the value when the population of the metastable level becomes important. Therefore, all computations here and below are made for case A of the recombination-photoionization of atomic hydrogen. We supported this conclusion by computations of the populations for hydrogen levels using the publicly available code Cloudy \citep{Ferland2017}: the values of hydrogen ionization fractions obtained by Cloudy and our code agree well. 

We consider here the simple evolution model of the radiation temperature of the first light:
\begin{equation}
T_{fl}=T_*\left(\frac{1+z_{rei}}{1+z}\right)^p,\quad \mbox{\rm with}\quad p=n\left(\frac{1+z}{1+z_{rei}}\right)^{1/3},
\label{Tfl}
\end{equation}
with such values of parameters: 
a) $T_*=5000$ K, $n=2$ (Ia), b) $T_*=10000$ K, $n=2$, (Ib), c) $T_*=5000$ K, $n=1$  (Ic), and d) $T_*=10000$ K, $n=1$ (Id). So, at the moment of reionization, $z_{rei}$, the radiation temperature of the first light is 5000 K or 10000 K. For the dilution factor, $\alpha_{fl}$, we set the constant and estimate its value using the condition of reionization. The condition $x_{HI}=x_{HII}=0.5$ at $6\le z_{rei}\le8$, which we suppose here, gives $3.3\cdot10^{-10}\le\alpha_{fl}\le4.85\cdot10^{-10}$ for model Ia, 
$6.0\cdot10^{-17}\le\alpha_{fl}\le8.6\cdot10^{-17}$ for model Ib,
$1.9\cdot10^{-10}\le\alpha_{fl}\le3.0\cdot10^{-10}$ for Ic,
 and $4.5\cdot10^{-17}\le\alpha_{fl}\le6.3\cdot10^{-17}$  for Id.
So, for the same $T_*$ the range of values of $\alpha_{fl}$ is narrow for such a range of redshift of reionization, but the value strongly depends on $T_*$.  

The spectral energy densities for all these models are shown in Fig.~\ref{flI}. The potential of ionization of hydrogen atoms is in the Wien range of the Planck distribution. Thus, the higher the radiation temperature at the moment when $x_{HII}\approx0.5$, the lower the dilution coefficient should be.
For comparison, we present in this figure the spectral energy density of CMBR at the same redshifts as well as at the current epoch ($z=0$). The observable cosmic infrared background (CIB) \citep{Kogut2019} is shown as well. Such a comparison shows that at frequencies below $2.5\cdot10^4$ GHz ($\lambda>0.1\,\mu m$), the intensity of such a first light is essentially lower than CMBR and CIB. However, it can excite the ro-vibrational levels of the most abundant molecules to produce the spectral distortions of the CMBR in the shortwave range. 

In the left panel of Fig.~\ref{xHII}, we present the evolution of the hydrogen ionized fraction $x_{\rm{HII}}$ for the models of the first light Ia-Id. The helium atoms are represented by the neutral fraction. The values of dilution coefficients, $\alpha_r$, are adjusted so that $x_{\rm{HII}}\approx0.5$ on $z = 8$ and 6. We can see that the data in this figure well match the data in Fig.~45 of \cite{Planck2020}. It means that these models of the first light are in agreement with the observational data on the optical depth of reionization.

In the right panel of Fig.~\ref{xHII}, we present the ranges of dilution coefficient against the temperature of thermal radiation for two evolution models of the first light, which provide $x_{\rm HII}\approx0.5$ at $z_{rei}=8$ (solid line) and $z_{rei}=6$ (dashed line). 

\section{Molecules in the inter-proto-galactic medium}

\begin{figure*}[ht!]
\includegraphics[width=0.49\textwidth]{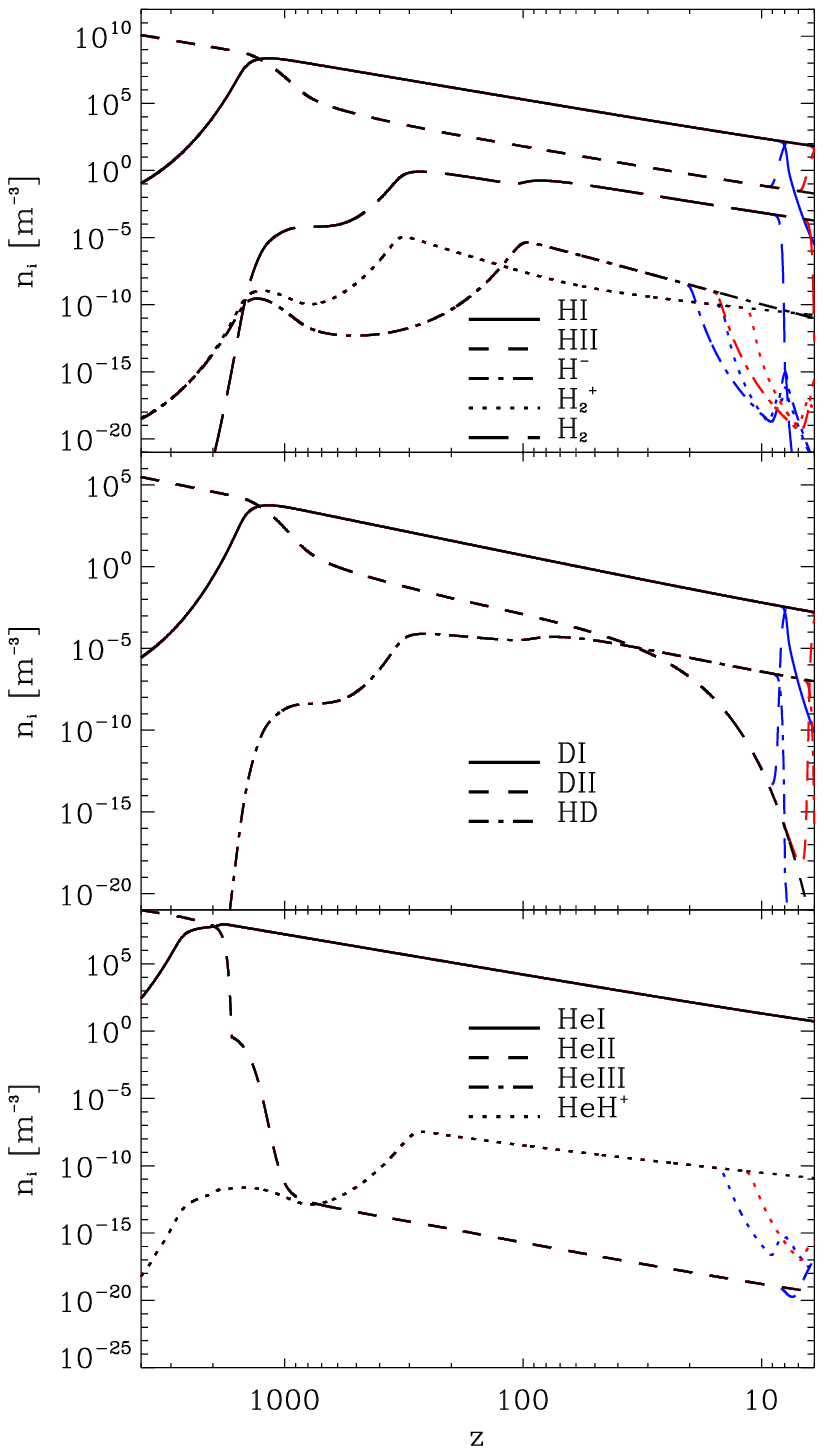}
\includegraphics[width=0.49\textwidth]{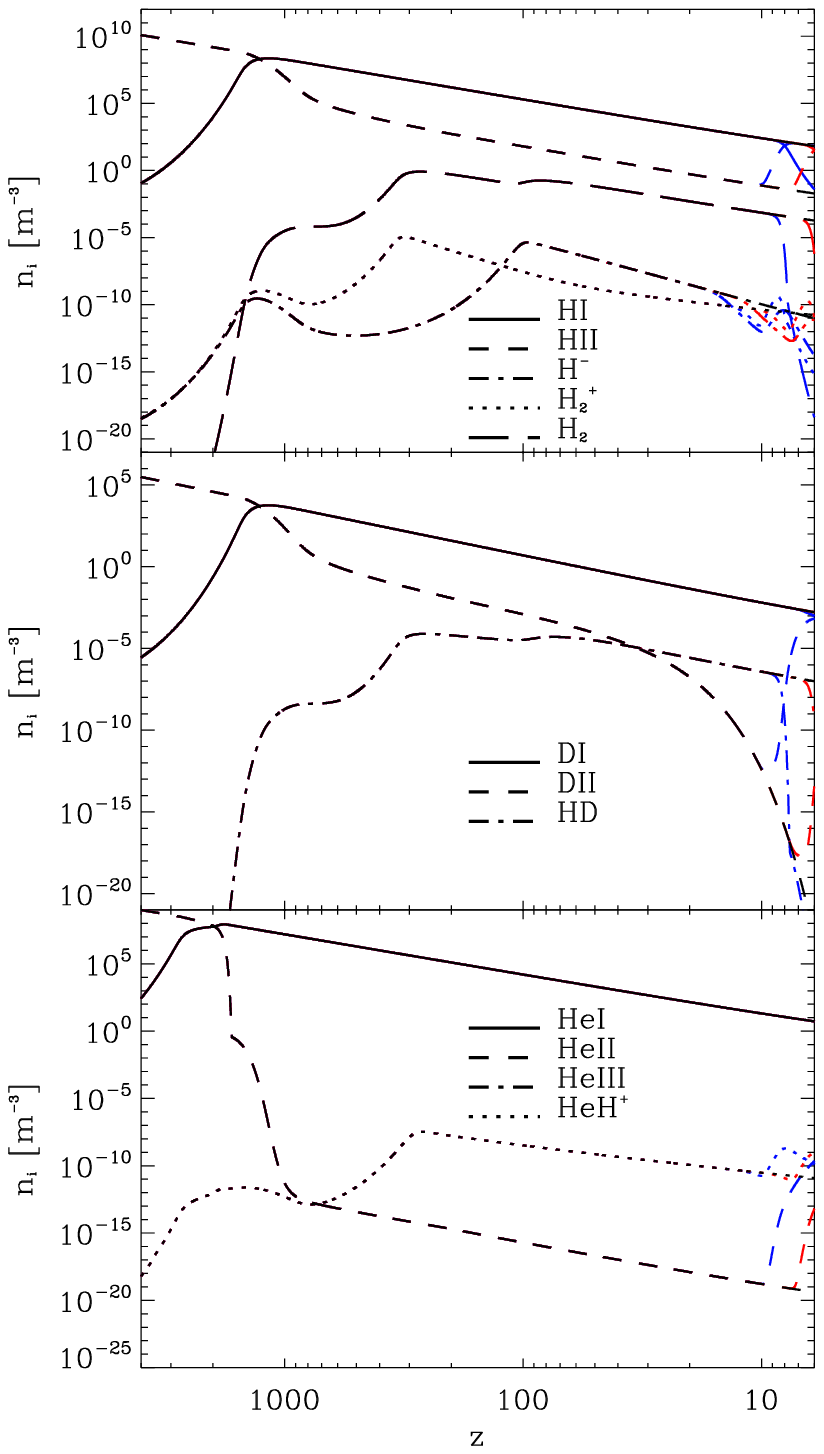}
\caption{Evolution of the number densities of atoms, ions, neutral, and ionized molecules $n_i(z)$ from cosmological recombination through the Dark Ages and Cosmic Dawn up to reionization for the models of the first light Ia and Ib: the red lines for  $z_{rei}=6$, the blue lines for $z_{rei}=8$. The black lines are the number density of species for the case without the first light, $n_i^0(z)$.}
\label{nIab}
\end{figure*}

The evolution of the number density H, D, and He and the temperature of CMBR and baryonic matter through the Dark Ages and Cosmic Dawn up to reionization are shown in Fig.~\ref{roT} (top panel and bottom one correspondingly). In the bottom panel, we also show the evolution of radiation temperature of the first light. The correspondence of the lines of the radiation temperature to the models of the first light is the same as in Fig.~\ref{flI}. The heating of the baryonic gas by the first light is noticeable. We can see that the dilution factor is important in the heating of gas by radiation: the thermal radiation with lower $T_{fl}$ but higher $\alpha_{fl}$ (model Ia and Ic) heats more than the thermal radiation with higher $T_{fl}$ but lower $\alpha_{fl}$ (models Ic and Id). The temperature of baryonic matter is approximately the same ($\sim2000$ K) in the cases of models Ia and Ic at the moment of reionization $z=8$ and 6, as it is shown in Fig.~\ref{roT} (solid red lines of $T_b$). At the same time, the temperature of the first light is 5000 K. For the model of the first light with radiation temperature $T_{fl}=10000$ K at the moment of reionization, the dilution factor is $\sim10^{-17}$ and the baryonic matter is heated by radiation only by a few degrees (solid blue lines of $T_b$).

\begin{figure*}[ht!]
\includegraphics[width=0.49\textwidth]{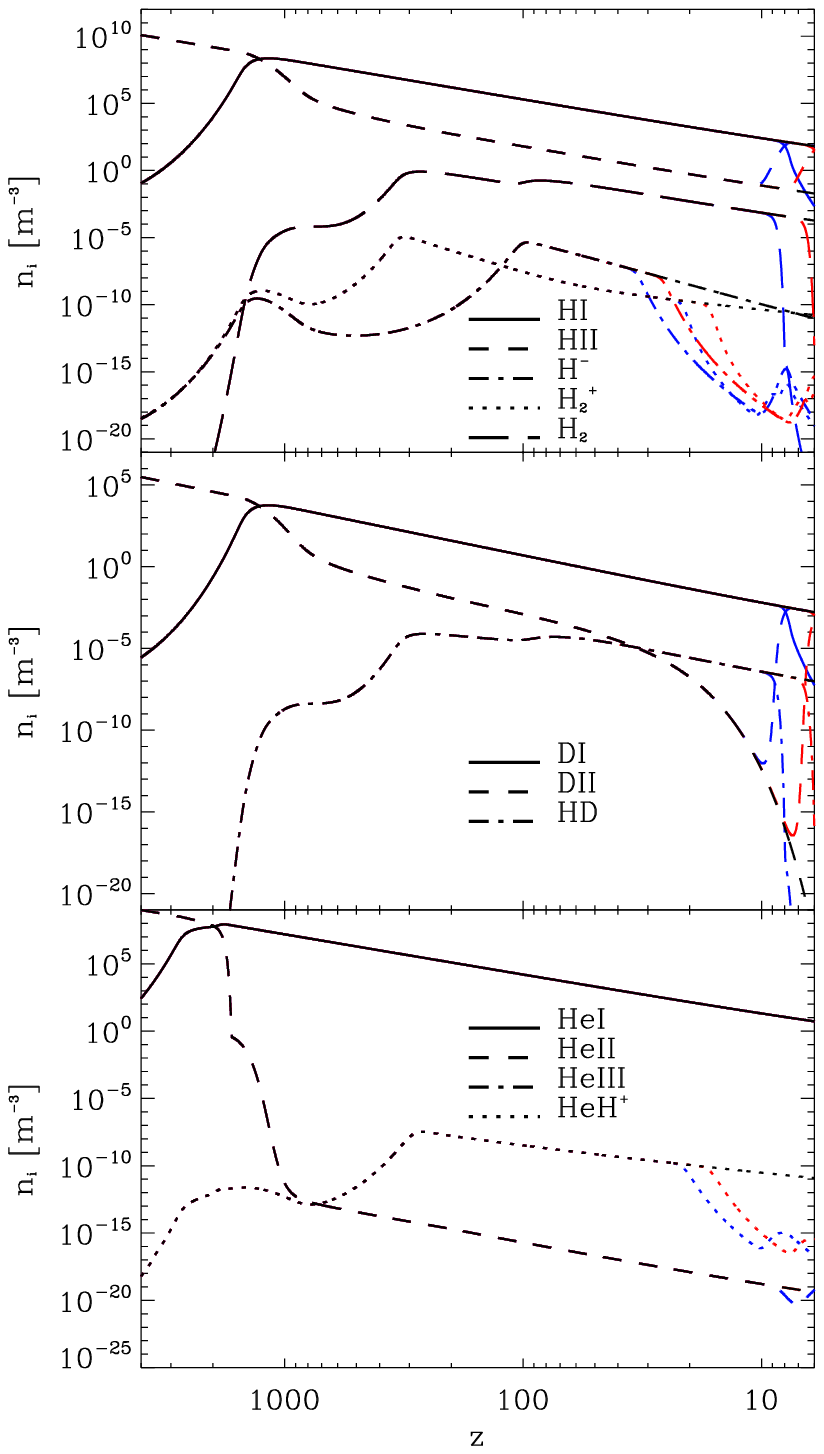}
\includegraphics[width=0.49\textwidth]{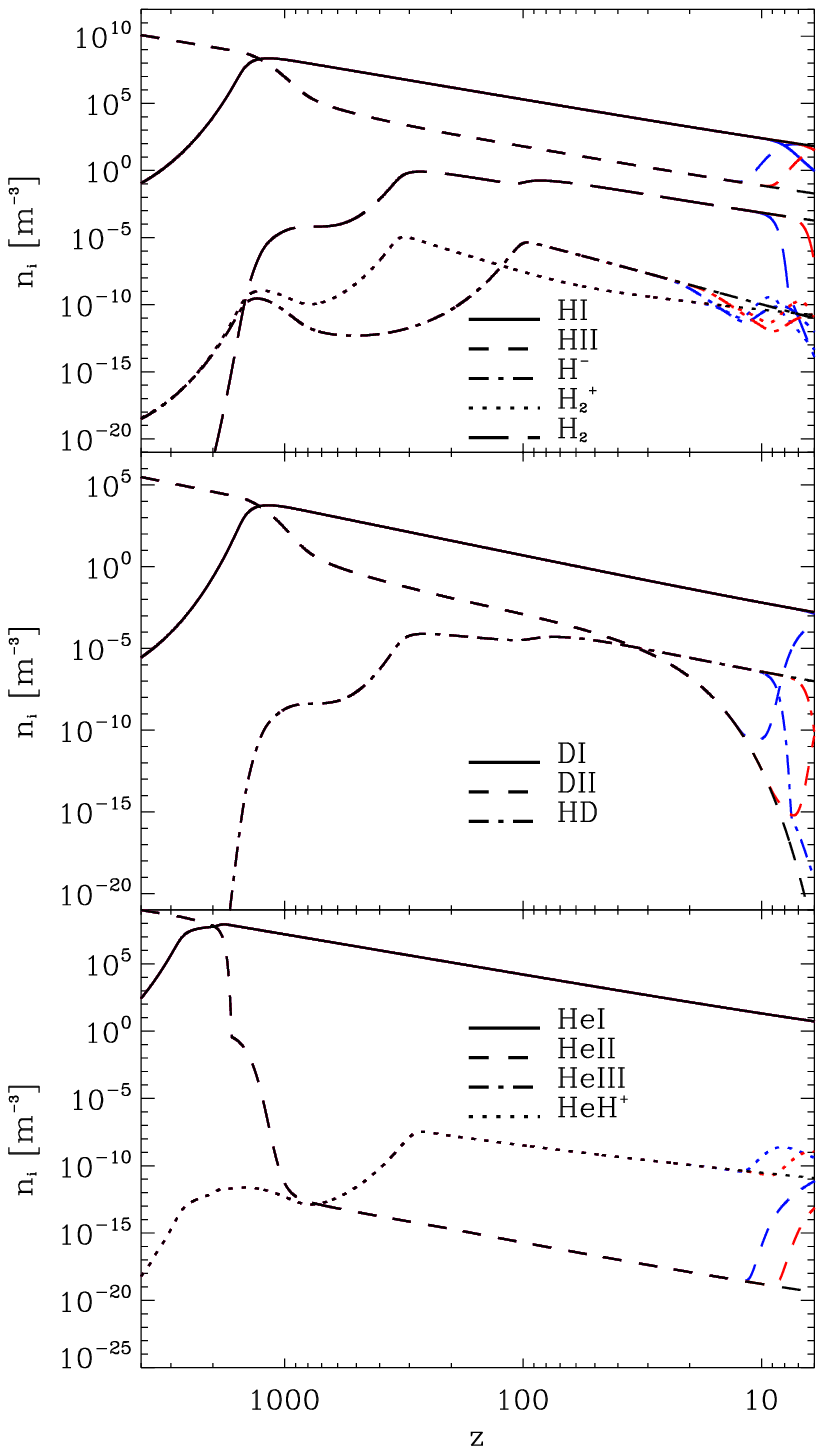}
\caption{Evolution of the number densities of atoms, ions, and neutral and ionized molecules $n_i(z)$ from cosmological recombination through the Dark Ages and Cosmic Dawn up to reionization for the models of the first light Ic and Id: the red lines for  $z_{rei}=6$, the blue lines for $z_{rei}=8$. The black lines are the number density of species for the case without the first light, $n_i^0(z)$. }
\label{nIcd}
\end{figure*}
\begin{figure*}[ht!]
\includegraphics[width=0.49\textwidth]{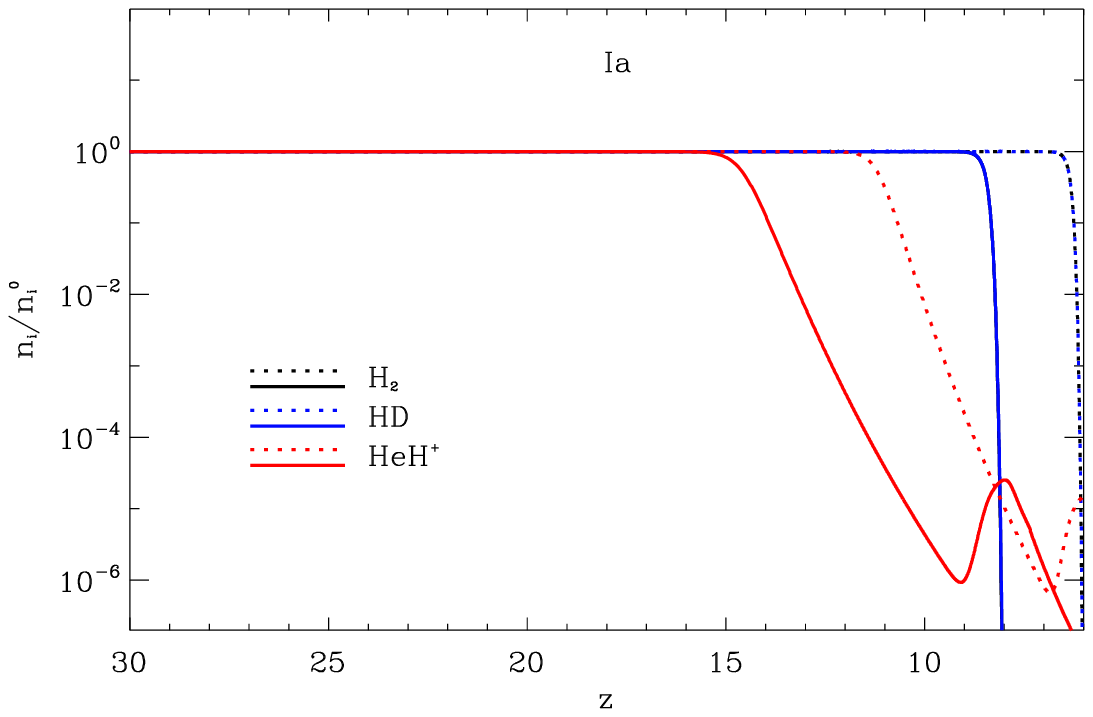}
\includegraphics[width=0.49\textwidth]{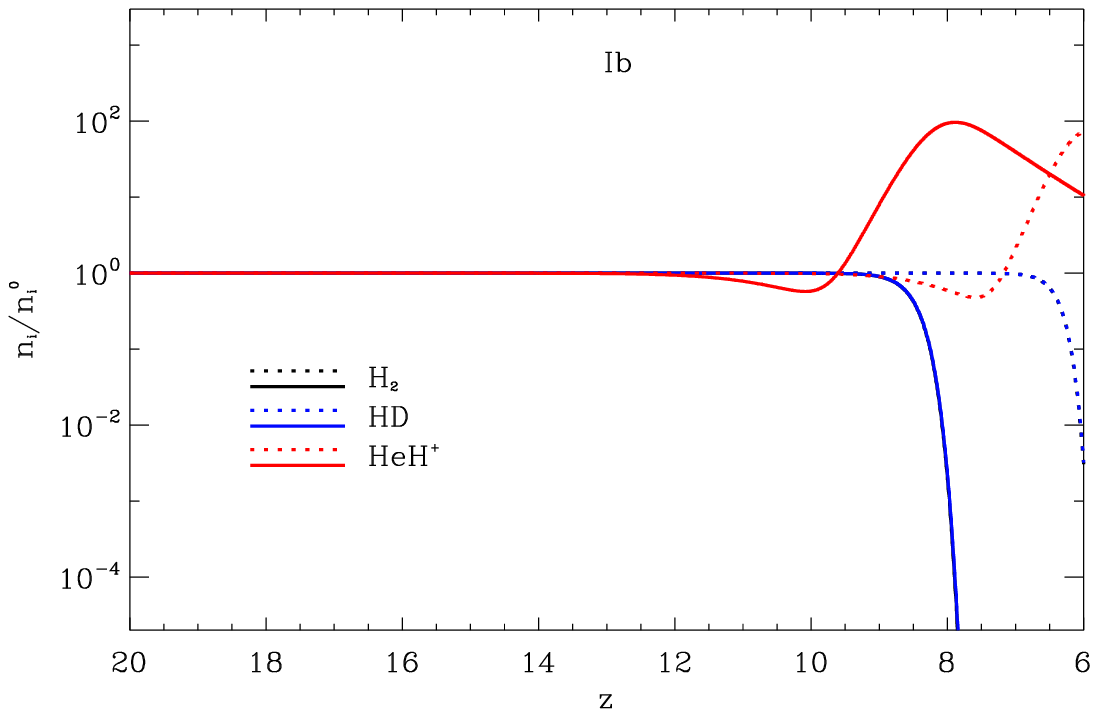}
\includegraphics[width=0.49\textwidth]{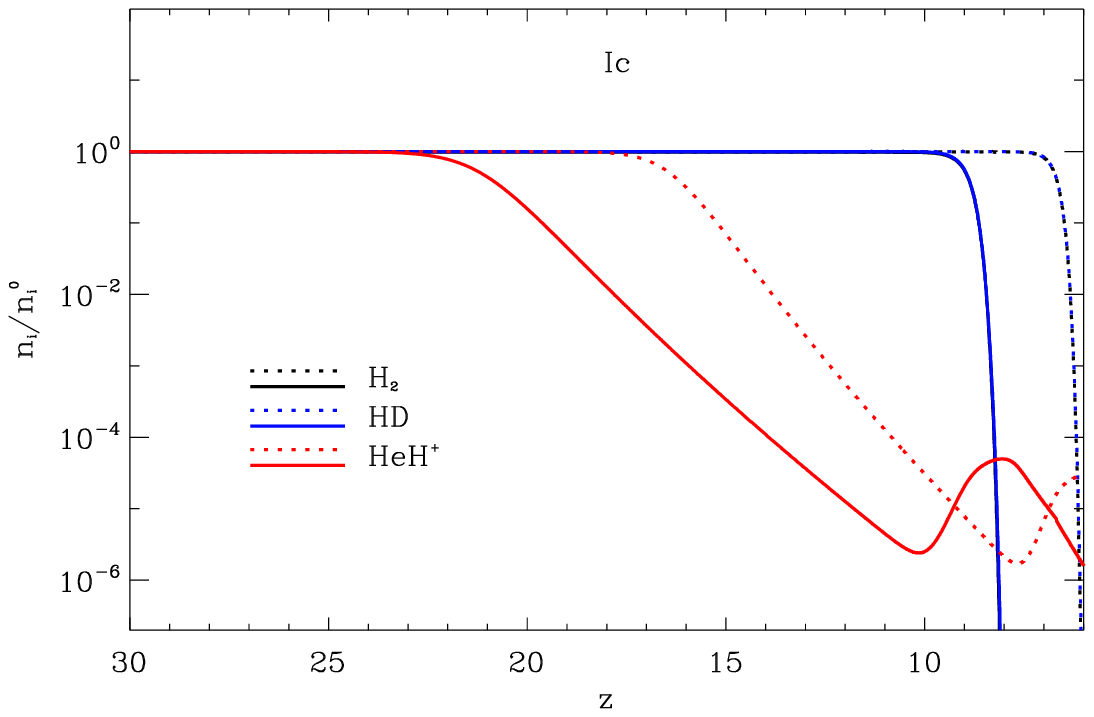}
\includegraphics[width=0.49\textwidth]{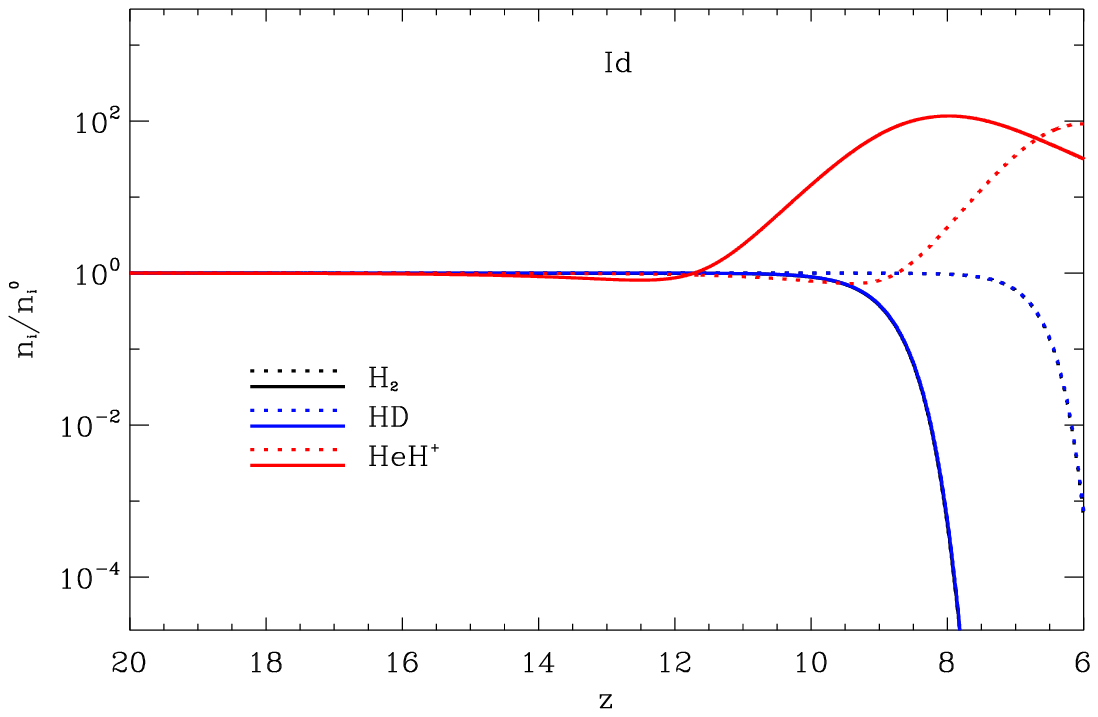}
\caption{Impact of the first light on the number densities of molecules at Cosmic Dawn for the different evolution of the energy density of thermal radiation of the first sources (models Ia-Id) (lines for H$_2$ and HD superimposed). }
\label{iIa-d} 
\end{figure*}
We computed the evolution of the number densities of atoms and molecules, $n_i$, starting from the early epoch before the cosmological recombination ($z=10000$) up to the beginning of the epoch of reionization ($z=6$) for different models of the first light. In Figs.~\ref{nIab} and \ref{nIcd}, we present the results for the Ia -- Id models of the first light. The black lines show the evolution of the number densities of atoms, ions, neutral and ionized molecules without the first light, $n_i^0$, which well agree with results by \cite{Galli1998,Schleicher2008}; and \cite{Vonlanthen2009}. Recent studies by \cite{Maio2022} of H$_2$ gas evolution over cosmic time ($2\le z\le17$) in the cosmological simulation ColdSIM do not contradict the semi-analytical ones presented here. The fact that the H$_2$ molecules at $z\le8-9$ do not disappear due to photodissociation can be explained by the effect of self-shielding of dense or extended regions of molecular hydrogen, which is taken into account in those studies. The lines of the same atoms and molecules in the epochs preceding the Cosmic Dawn are superimposed and black. The color lines show the influence of the first light on the number density of the first molecules. 

We can see that negative hydrogen ions $\rm{H}^{-}$ are most sensitive to the first light. This is expected since its threshold of dissociation is low, namely, 0.75 eV. The photodissociation reaction $\mathrm{H^- + \gamma\rightarrow H + e^-}$ effectively decreases the number density of $\rm{H}^{-}$ up to the moment when the ionization of hydrogen starts to increase. Increasing the number density of free electrons makes raises the importance of the reaction $\mathrm{H + e^- \rightarrow H^- + \gamma}$, which leads to the growth in the number density of  $\rm{H}^{-}$.  This is not for long, however, as the rapid increase of the number density of photons subsequently makes the process of photodissociation dominant. 

The evolution of the number densities of the molecular ions $\rm{H}_2^+$ and $\rm{HeH}^+$ is similar for the high energy density of the first light because their photodissociation thresholds (2.77 eV and 1.85 eV accordingly) are essentially lower than the photoionization potential of atomic hydrogen (13.6 eV). In both cases, an increase in the value of $n_{\rm{HII}}$ causes a brief increase in the number density of these molecular ions, as well as a subsequent decrease due to the growth of the photodissociation. For the low energy density of the first light (bottom panels in the right columns of Figs.~\ref{nIab} and \ref{nIcd}), the number density of hydride helium ions $HeH^+$ increases when the ionization of atomic hydrogen increases because the reaction $\mathrm{He + H^+ \rightarrow HeH^+ + \gamma}$ prevails over the reaction $\mathrm{HeH^+ + \gamma \rightarrow He + H^+}$. 

The number density of the most abundant molecules $\rm{H}_2$ (threshold of dissociation is 4.53 eV) drastically decreases when the number density of photons with energy above the threshold becomes larger than a specific value when the rate of direct dissociation of $\rm{H}_2$ becomes $\sim10^{-18}-10^{-19}$ $\mathrm{cm^3s^{-1}}$.
Thus, the spectral energy distribution of the first light and its time dependence crucially impact the formation and dissociation of molecules during the Cosmic Dawn. To demonstrate this impact more clearly, we  present in Fig.~\ref{iIa-d} the ratios of the number densities of molecules in the models with the first light, $n_i(z)$ to the number density of molecules without it, $n_i^0(z)$.

In this study, we supposed the primordial chemical content of hydrogen, helium, and deuterium. Taking into account the PopII stellar nucleosynthesis,  metal enrichment will change the relationship between them, which will somewhat affect the number densities of the first molecules in the halos. Obviously, these changes will not be decisive at $z\ge10$.

\section{Molecules in the intra-proto-galactic medium and their luminescence}

We went on to repeat the same computations for the intra-proto-galactic medium. We supposed that the proto-galaxies are spherical top-hat halos in the multicomponent Universe, which are filled with dynamical dark energy, dark matter, baryon matter, and two ``sorts'' of thermal radiation, namely, the CMBR and the first light, which appeared in the Cosmic Dawn epoch. Since the time dependence of the density and temperature of the baryonic matter in the halos at stages before, during, and after virialization differ from the background ones, the number densities of the first molecules are different too. Moreover, the warm and hot halos can slightly shine at the cosmic microwave background \citep{Novosyadlyj2020,Kulinich2020}. To analyze the impact of the first light on the abundance of the first molecules and their luminescence in the halos, we use the theory of cosmological perturbations and the halo model to produce accurate computations of physical conditions and integrations of the corresponding kinetics equations.

\subsection{Dynamical model of halo}

\begin{figure}[ht!]
\includegraphics[width=0.49\textwidth]{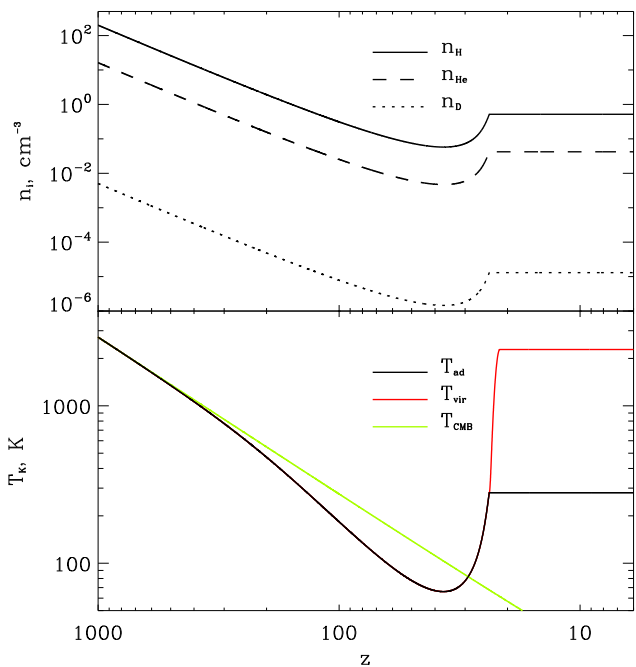}
\caption{Formation of a spherical halo from a specific cosmological perturbation. Top panel: Evolution of number densities of H, D, and He in the top-hat halo with $k$=80 Mpc ($M_h=1.26\cdot10^{-6}$ M$_\odot$) and initial amplitude of adiabatic scalar perturbation $\delta_r=1.5\cdot10^{-4}$ at $z=10^8$). Bottom panel: Evolution of the temperature of CMBR (green line) and kinetic temperature of the top-hat halo with adiabatic heating (dark line) and virial temperature (\ref{Tvir}; red line).}
\label{roT_halo}
\end{figure}

We computed the physical conditions and chemistry of the halo by modeling the evolution of a single spherical perturbation in the four-component Universe (cold dark matter, baryon matter, dark energy, and thermal relict radiation) starting from the linear stage at the early epoch, long before the cosmological recombination, through the quasi-linear stage, turnaround point and collapse up to virialized state \citep{Novosyadlyj2016,Novosyadlyj2018}. We assume that mass of a halo  $M_h\sim10^6$ M$_\odot$, which is the typical one in the cosmological model with cold dark matter and dark energy. The initial amplitudes of adiabatic perturbation in the conformal Newtonian gauge for this scale are taken as follows: for thermal radiation $\delta_{tr}=1.5\cdot10^{-4}$, for the matter components $\delta_{cdm}=\delta_b=1.125\cdot10^{-4}$, and for phantom dark energy $\delta_{de}=-3.376\cdot10^{-6}$ at $z_{init}=10^8$, that correspond to the root mean square values at this scale in the cosmological model with parameters presented in the Introduction. At the redshift $z_{init}$, such a perturbation is the superhorizon one, which enters the particle horizon at $z_{eh}\approx2.5\cdot10^6$. The integration of the equation system that follows from those of Einstein and energy conservation \citep{Novosyadlyj2016,Novosyadlyj2018} gives the moment of virialization $z_v=23.8$ when the overdensity $\rho_m(z_v)/\bar{\rho}_m(z_v)$ reaches the value $\Delta_v=178$ (top panel of Fig.~\ref{roT_halo}).

The halo mass $M_h$ [$M_{\odot}$], the comoving wave number of linear (seed) perturbation $k$ [Mpc$^{-1}$]  and the comoving radius $r_h$ [kpc] after virialization are related as:
\begin{equation}
\frac{M_h}{M_{\odot}}\approx1160\Delta_v(1+z_v)^3\Omega_mh^2r_h^3\approx4.5\cdot10^{12}\Omega_mh^2k^{-3},\nonumber
\end{equation}
where $\Omega_m\equiv8\pi G\bar{\rho}_m(0)/3H_0^2=\Omega_{dm}+\Omega_b$ and $h\equiv H_0/100\,\rm{km/s/Mpc}$. In the cosmological model with $\Omega_m=0.315$ and $h=0.674$ for $k=80$~Mpc$^{-1}$, we obtain the value $M_h\approx1.26\cdot10^6$ M$_\odot$ and $r_h\approx0.15$~kpc. The angular size of such halo at $z_v\div z_{rei}$ is in the range of $0.07\div0.03$ arcsec. We suppose that a halo is spherical and homogeneous (top-hat) with values of matter density and kinetic temperature unchanged after virialization (Fig.~\ref{roT_halo}). The total number density of baryon particles is unchanged too and equals 0.68 cm$^{-3}$.

The kinetic temperature of baryon gas in the halo during the epochs of the Dark Ages and Cosmic Dawn is determined by its interaction with thermal relic radiation and the first light and the adiabatic heating  over compression. The bremsstrahlung, molecular cooling and other cooling mechanisms are not effective due to the low gas density (top panels of Fig.~\ref{roT} and \ref{roT_halo}). So, the evolution of baryon gas temperature $T_b$ at $6\le z\le850$ can be obtained by the integration of the energy balance equation for the intra-halo baryon matter:
\begin{eqnarray}
\frac{dT_{b}}{dz}=\mbox{r.h.s.}\,(\ref{Tb})+\frac{2}{3}\frac{T_{b}}{1+z}\frac{d\delta_b}{dz}
\label{Tad} 
,\end{eqnarray}
The kinetic temperature of the gas in such halo after virialization is about 280 K and does not depend on halo mass. The halo with such temperature we call the "warm"\ halo.

The gas in the halo can also be heated by shocks in the processes of violent relaxation to the virial temperature \citep{Barkana2001,Bromm2011}:
\begin{equation}
T^{\rm (vir)}_b=2\cdot10^4\left(\frac{\mu_{\rm H}}{1.2}\right)\left(\frac{M_h}{10^8\,\mathrm{M_{\odot}}}\right)^{2/3}\left(\frac{\Delta_v}{178}\right)^{1/3}\left(\frac{z_v+1}{10}\right)\,\mathrm{K},
\label{Tvir}     
\end{equation}
where $\mu_{\rm H}$ is the mass per H atom. In our case, the temperature of such halo is 2280 K and we call it the  "hot"\ halo.

In the bottom panel of Fig.~\ref{roT_halo}, the temperature of the gas in the virialized halo is shown for two cases: when it is heated by adiabatic compression only (dark solid line) and when it is heated in the processes of violent relaxation to virial temperature (\ref{Tvir}; red solid line). In the first case, the final temperature is a result of the integration of Eq. (\ref{Tad}), in the second one, we set it equal to $T^{\rm (vir)}_{\rm b}$ by hand at $z\ge z_v$.

\subsection{Number density of molecules}

In the halo model described above, the number density of baryon (and lepton) particles is unchanged after virialization, while the number densities of molecules, protons ($n_p$) and electrons ($n_e$) are  changed via recombination and photoionization processes of atoms and molecules as well as  the formation and destruction reactions of molecules. 

\begin{figure}[ht!]
\includegraphics[width=0.49\textwidth]{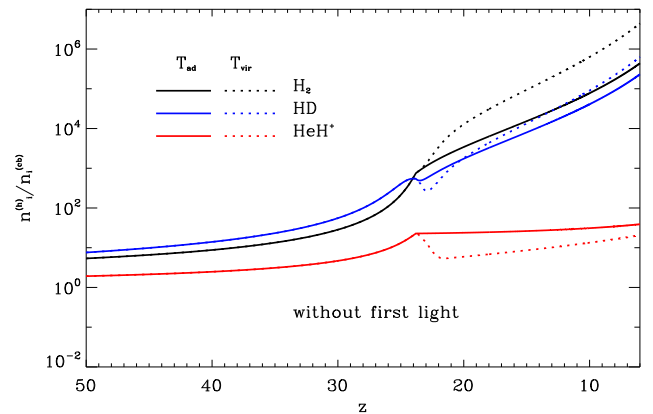}
\caption{Ratio of the number densities of molecules in the warm and hot halo ($n^{(h)}_i$) to the number density of molecules at the cosmological background ($n^{(cb)}_i$) without the first light.}
\label{whhn2bn}
\end{figure}
In Fig.~\ref{whhn2bn}, we show how the number densities of molecules H$_2$, HD, and HeH$^+$ are changed in the warm and hot halos in comparison with the background ones (inter-proto-galaxy medium) without the first light. We can see that increasing of density and temperature of baryon gas leads to the increasing of the number density of molecules H$_2$ and HD even after virialization essentially more than $\Delta_v$ times. Moreover, in the hot halos, the number density of these molecules is higher than in the warm ones. The number density of HeH$^+$ molecular ions behaves differently. In the warm halo, the $n_{HeH^+}$ decreases so that the ratio $n_{HeH^+}/n^{0}_{HeH^+}$ is almost constant. In the hot halo, the number density of these molecules sharply decreases since destruction reactions He11 and He12 (Table 1) dominate.

\begin{figure*}[ht!]
\centering
\includegraphics[width=0.49\textwidth]{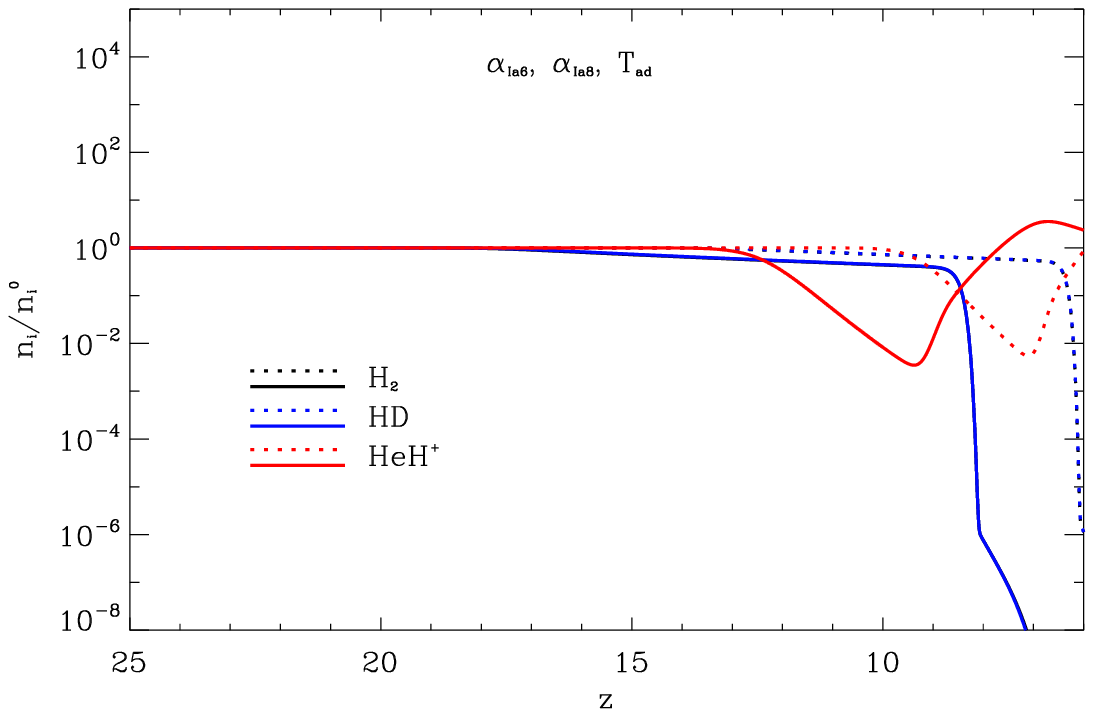}
\includegraphics[width=0.49\textwidth]{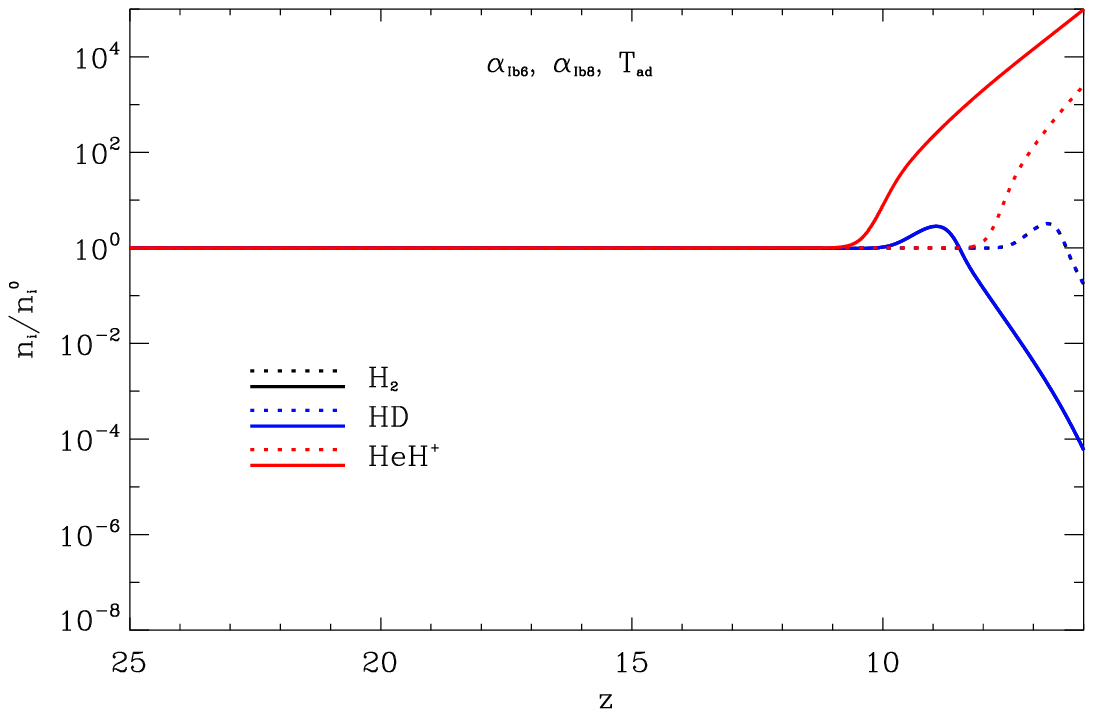}
\includegraphics[width=0.49\textwidth]{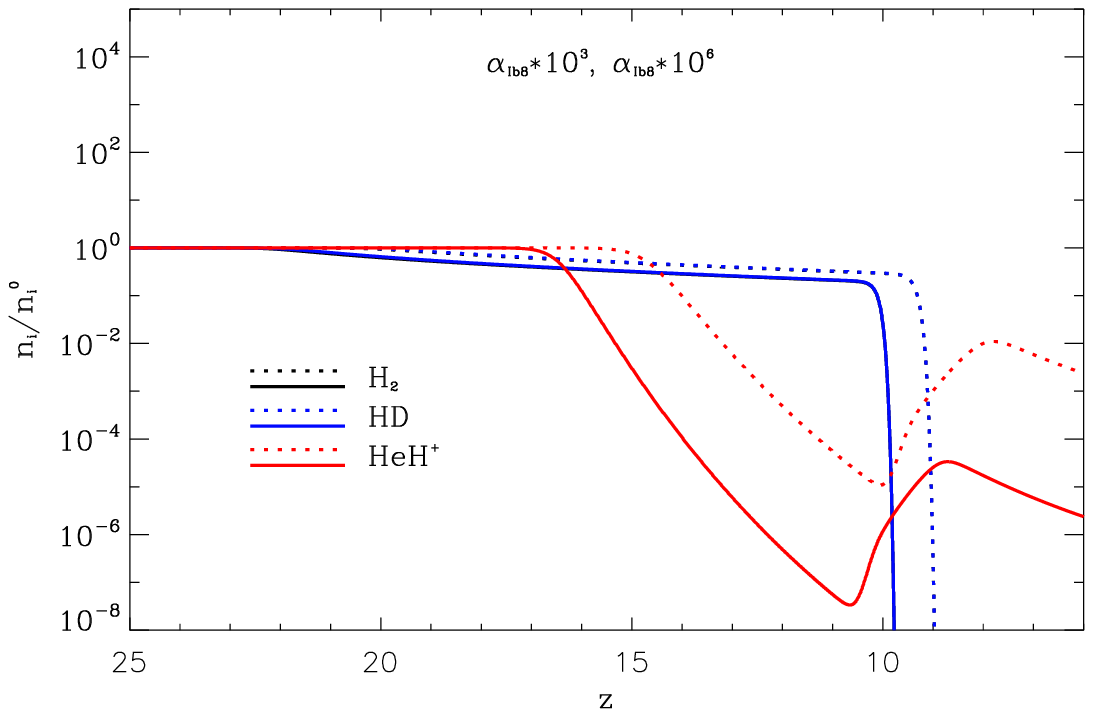}
\includegraphics[width=0.49\textwidth]{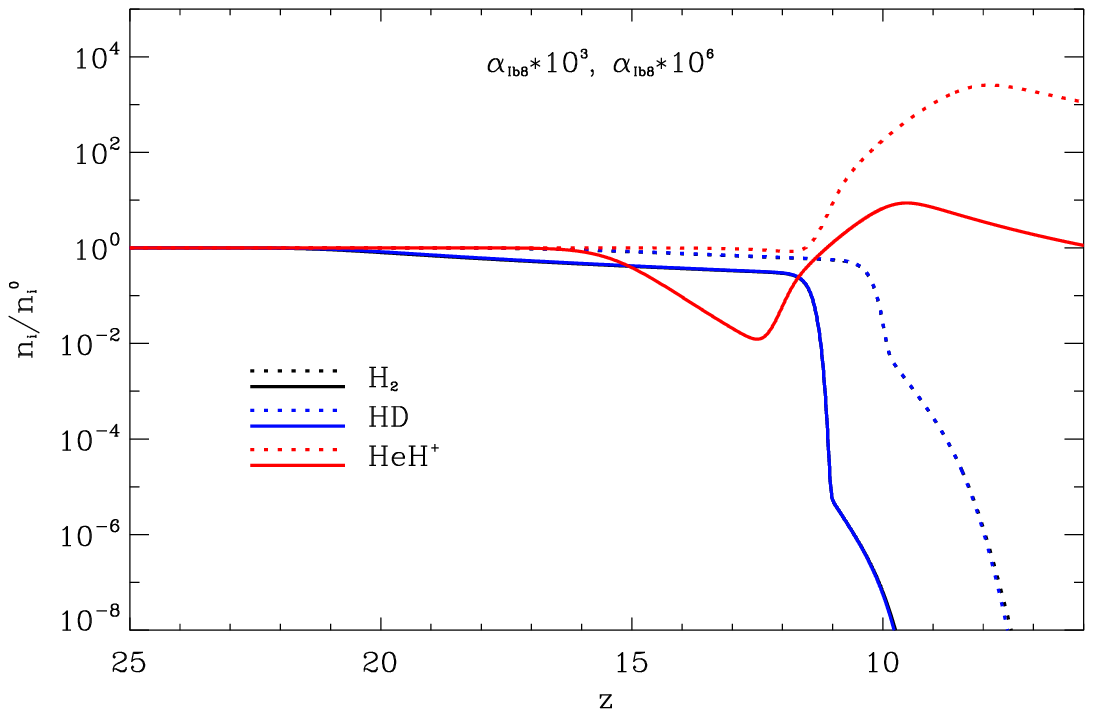}
\caption{Impact of the first light on the number densities of molecules in the halos of Cosmic Dawn. In the top row, the models of the first light are the same as in background ones (left - Ia, right - Ib), which reionize the medium at $z$=6 (dotted lines) and $z$=8 (solid lines). In the bottom row, the energy density of the first light in the halos are $10^3$ (dotted line) and $10^6$ (solid lines) times higher than energy density in the corresponding models in the top panels shown by solid lines. In all the panels, the lines for H$_2$ and HD are superimposed.}
\label{h_Ia-Ib}
\end{figure*}

To estimate the impact of the first light on the chemistry of the intra-halo medium, it is necessary to note that the spectral density of radiation in the halo is larger than the background one and this excess increases over time quickly during the Cosmic Dawn epoch. We analyzed the impact of the first light with different values of dilution coefficients on the halo chemistry: i) $\alpha_{fl}$, estimated in Section 3; ii) $10^3\alpha_{fl}$; and iii)$10^6\alpha_{fl}$. We computed the ratios of the number densities of molecules in the halo with the first light to the number densities of molecules in the halo without it.  The results for the first light models Ia and Ib are presented in Fig.~\ref{h_Ia-Ib}. In the top panels the models of the first light in the halo are the same, which reionize the inter-halo medium at $z$=6 (dotted lines) and $z$=8 (solid lines). In the left panel, the radiation temperature of the first light at the moment of reionization is 5000 K, with dilution coefficients $\alpha_{Ia6}=3.3\cdot10^{-10}$ and  $\alpha_{Ia8}=4.85\cdot10^{-10}$ for $z_{rei}=6$ and 8, correspondingly. In the right panel, the radiation temperature of the first light at the moment of reionization is 10000 K, with dilution coefficients $\alpha_{Ib6}=6.0\cdot10^{-17}$ and  $\alpha_{Ib8}=8.6\cdot10^{-17}$ for $z_{rei}=6$ and 8, correspondingly. In the bottom panels the energy densities of the first light in the halos are $10^3$ (dotted line) and $10^6$ (solid lines) times higher than the energy densities in the corresponding models in the top panels shown by solid lines. In all the panels, the lines for H$_2$ and HD are superimposed. We can see that these molecules are quite destroyed before reionization. The larger is the spectral density of the first light, the earlier they disappear because of photodissociation. 
 
The number density of HeH$^+$ molecular ions in the halo evolves quite differently. It starts to change early over the low threshold of dissociation and changes nonmonotonically over the competition of formation and destruction reactions (He8, He10, He11, He12, He14, and He15 in Table 1) in the medium with the growth of $n_{HII}$, $n_e$, and $n_\gamma$. Moreover, in the models of the first light with $\alpha\le10^{-14}$ and $T_*\sim10000$ K, the number density of these molecules increases by $10^2-10^3$ times on $z_{rei}$, when other molecules are destroyed in their entirety. It gives us the chance to observe the epoch of reionization in the lines of helium hydride ions. 

In order not to overload the paper we do not present the figures for models of the first light Ic and Id, in which the spectral energy distributions on $z_{rei}$  are the same as in the models Ia and Ib but with slower growth of temperature and energy density rate. As in the case of the inter-halo medium, the qualitative behavior is similar to the cases with Ia and Ib models, but the effect of the first light in the models Ic and Id is already noticeable at larger redshifts because its energy density is higher there.  

\subsection{Luminescence of halo in the molecular lines}

We went on to analyze the impact of the first light on the luminescence of a halo in the lines of the first molecules. Since the first light with temperature, $T_{fl}\sim10^3-10^{4}$, can excite the rotational-vibrational energy levels of the first molecules the luminescence of halos in the Cosmic Dawn would complicatedly depend on the spectral energy density of background radiation and the number density of the molecules. 

To evaluate the luminescence of the halo in the molecular lines, we must compute the population of rotational levels. We integrate for that the kinetic equations of population or depopulation of rotational levels: 
\begin{equation}
\frac{dn_i}{dt} = \sum\limits_{j\ne i}n_jR_{ji} - n_i\sum\limits_{j\ne i}R_{ij}+3Hn_i
\label{pop}
,\end{equation}
where the indices, $i,$ and $j,$ mark the rotational levels, $n_i$ is the number density of molecules at the rotational level $i$,  and  
\begin{equation}
R_{ij}=A_{ij}+B_{ij}U_{\nu_{ij}}+C_{ij}, \quad R_{ji}=B_{ji}U_{\nu_{ij}}+C_{ji} \nonumber
\end{equation}
are the rates of transitions between them. Here, $i>j$, $A_{ij}$ is the rate of spontaneous transition from $i$-level to $j$-level, $B_{ij}$ and $B_{ji}$ are the rates of induced transition and absorption at frequency $\nu_{ij}$, accordingly, $C_{ji}$ and $C_{ij}$ are the rates of collisional excitation of $i$-level from $j$-level and de-excitation from the $i$-level to the $j$-level, $U_{\nu_{ij}}=U_{\nu_{ij}}^{CMB}+U_{\nu_{ij}}^{fl}$ is the total energy density of radiation at any redshift from cosmological recombination to reionization. We used the same approximations for excitation and de-excitation rates of rotational levels of molecules H$_2$, HD, and HeH$^+$ as in our previous papers, \cite{Novosyadlyj2020} and \cite{Kulinich2020}, and references therein. 

\begin{figure*}[ht!]
\includegraphics[width=0.49\textwidth]{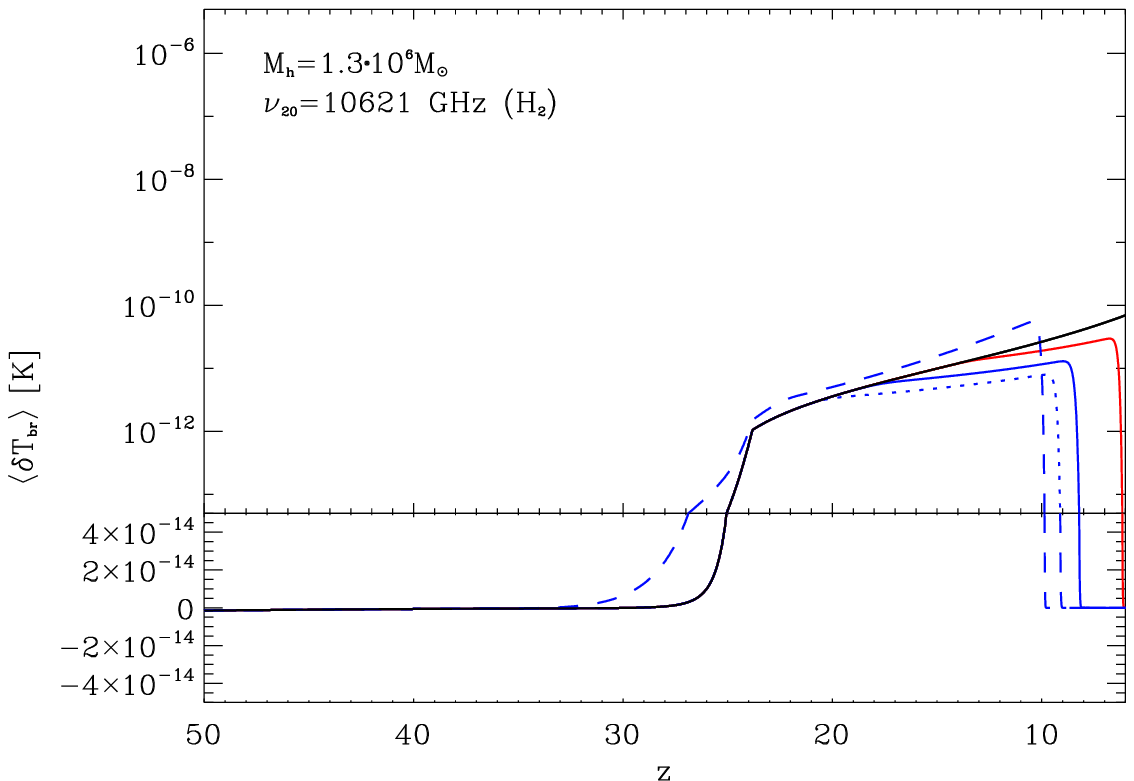}
\includegraphics[width=0.49\textwidth]{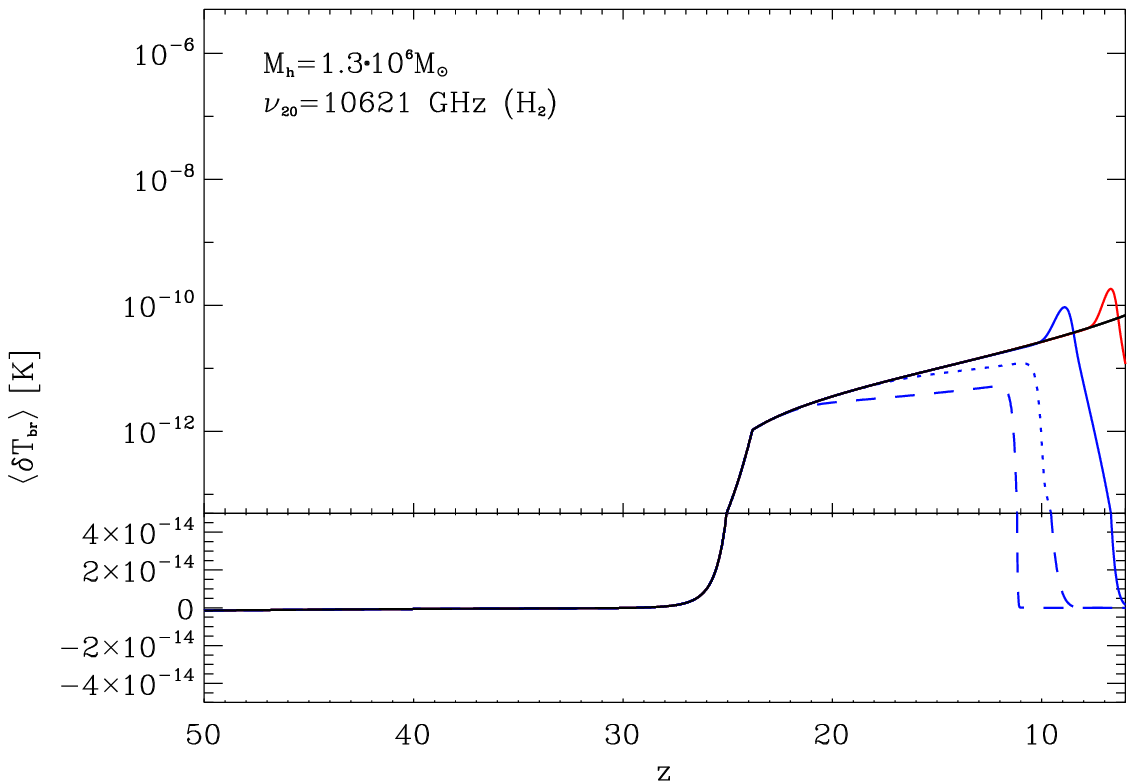}
\includegraphics[width=0.49\textwidth]{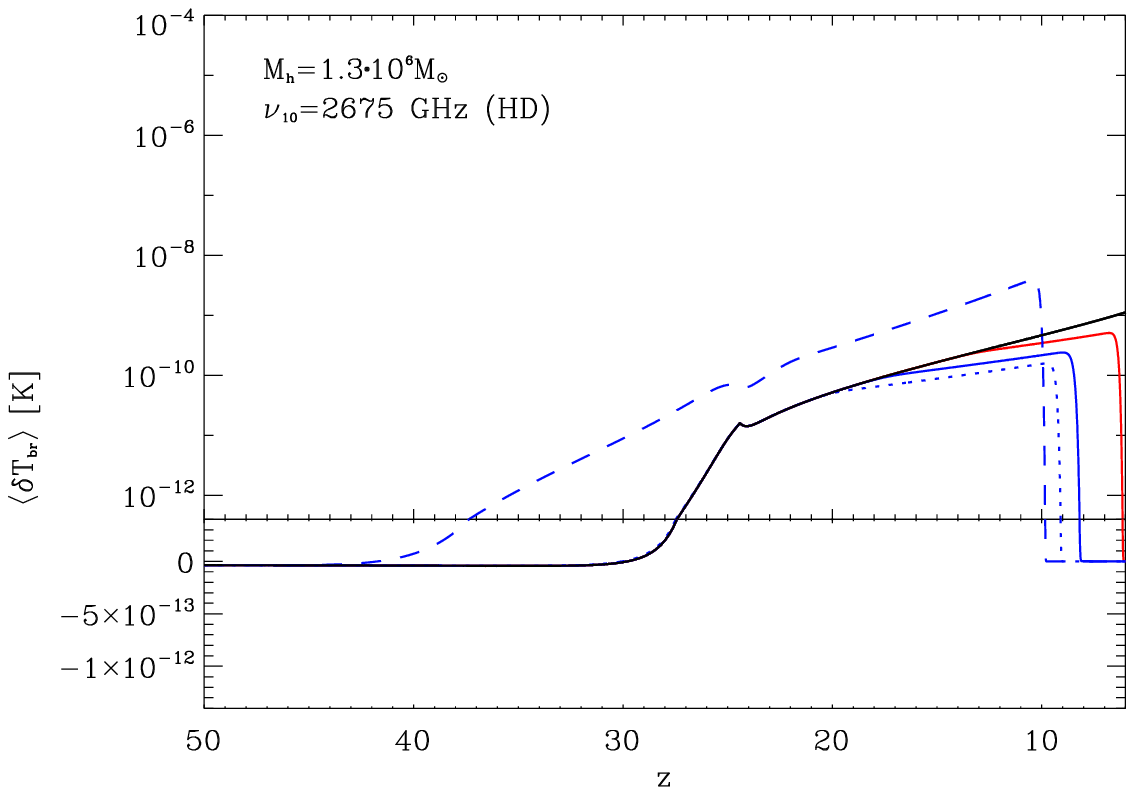}
\includegraphics[width=0.49\textwidth]{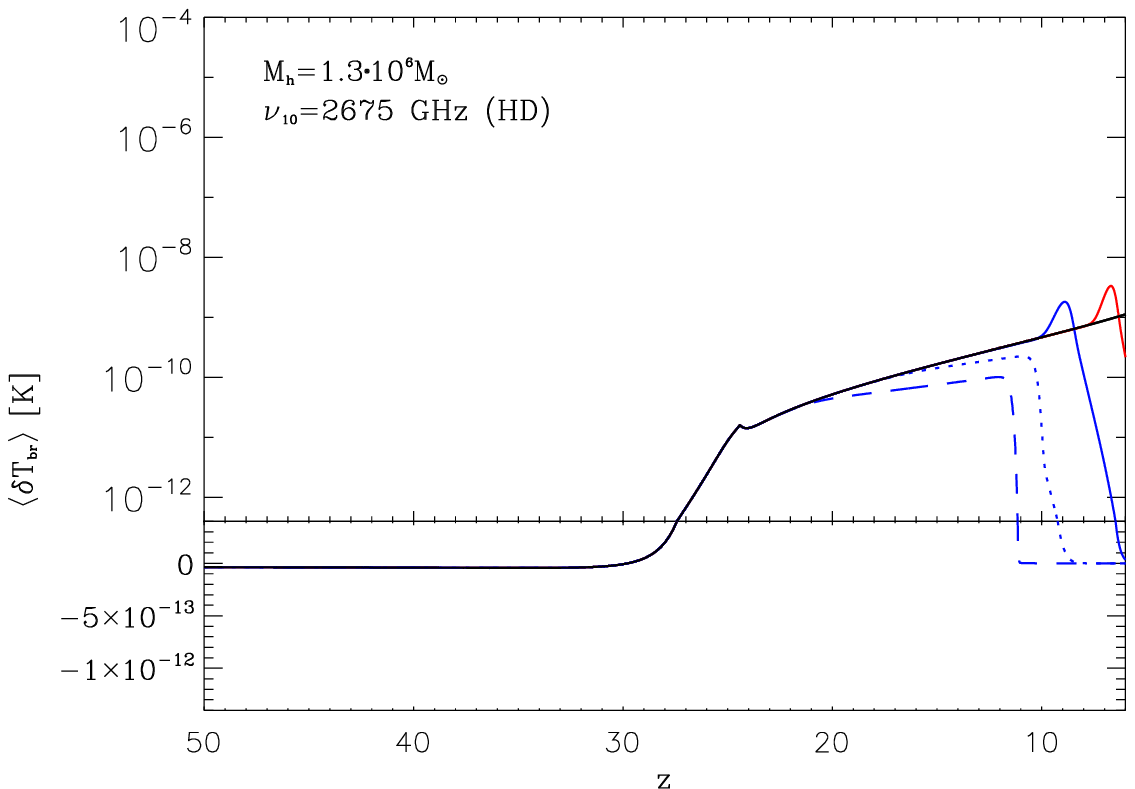}
\includegraphics[width=0.49\textwidth]{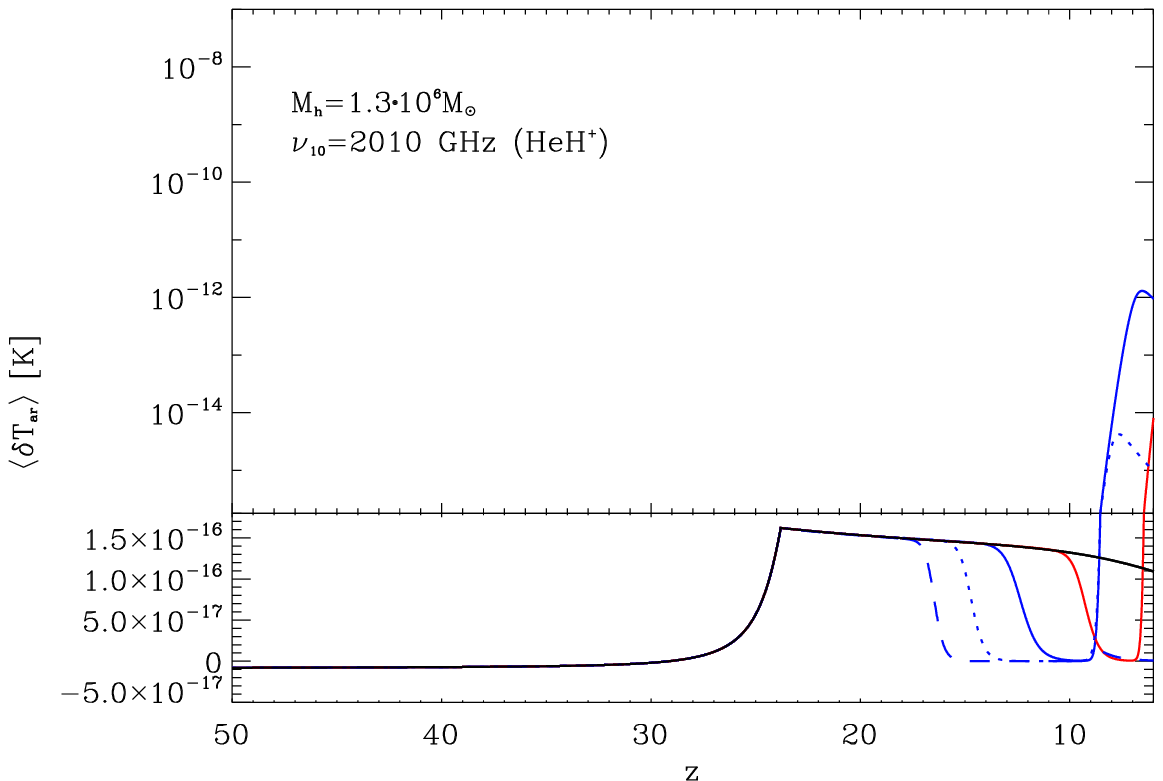}
\includegraphics[width=0.49\textwidth]{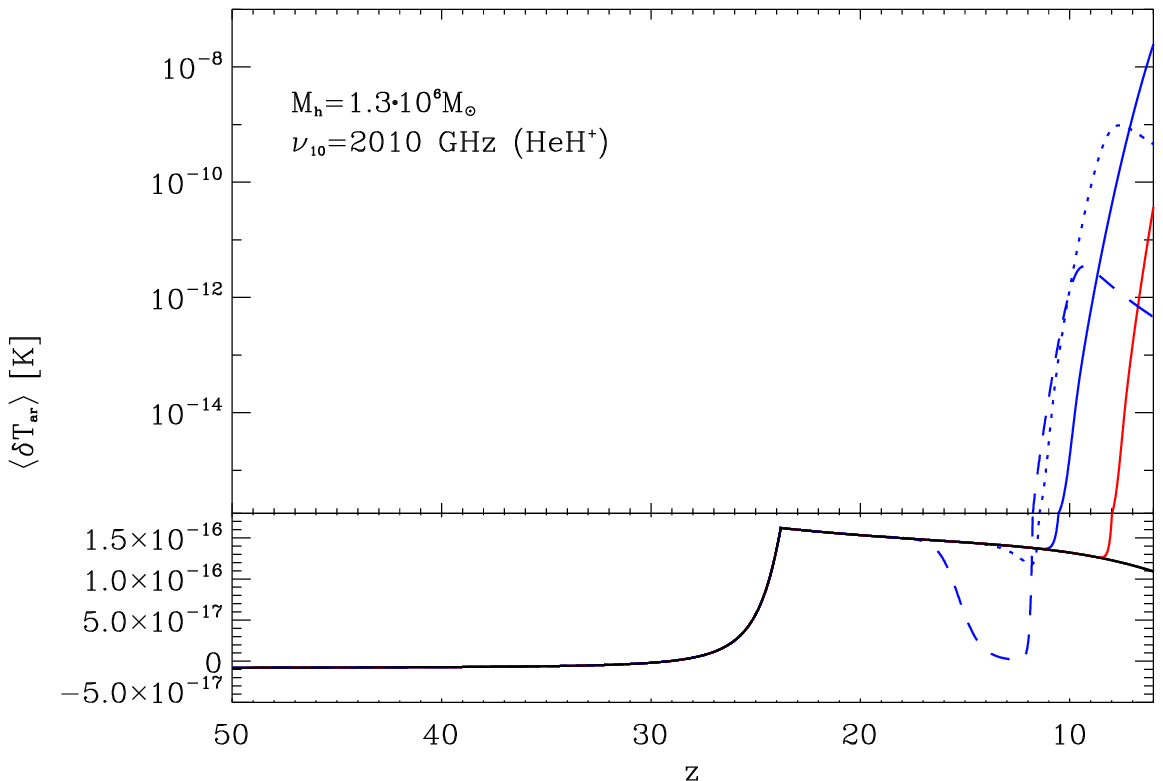}
\caption{Differential brightness temperature of warm halos ($T_b\approx280$ K) in the base rotational lines of molecules H$_2$, HD, and HeH$^+$ (from top to down) with and without of the first light: solid dark line -- without first light, solid red and blue lines with background first light which reionize the medium at z=6 and z=8 accordingly, dotted and dashed blue lines with the first light with energy density $10^3$ and $10^6$ times higher than the background one ($z_{rei}=8$). In the left column, the first light is Ia, in the right one it is Ib.}
\label{Tbr-wh}
\end{figure*}

\begin{figure*}[ht!]
\includegraphics[width=0.49\textwidth]{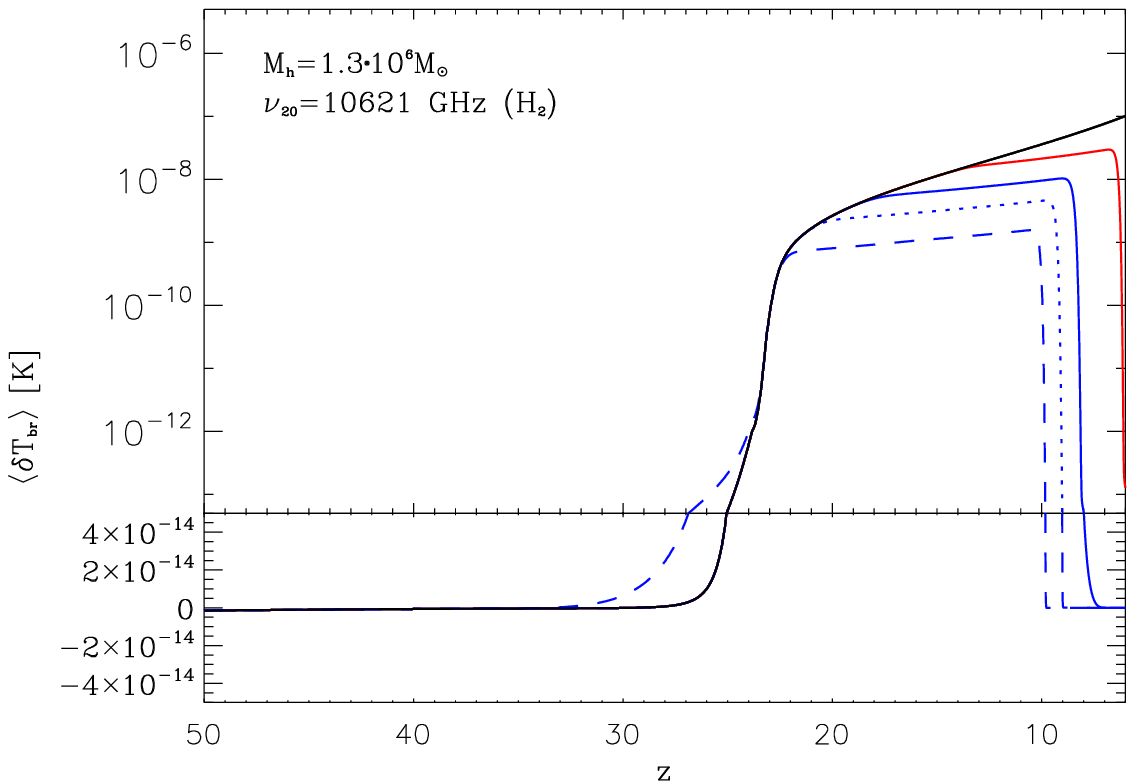}
\includegraphics[width=0.49\textwidth]{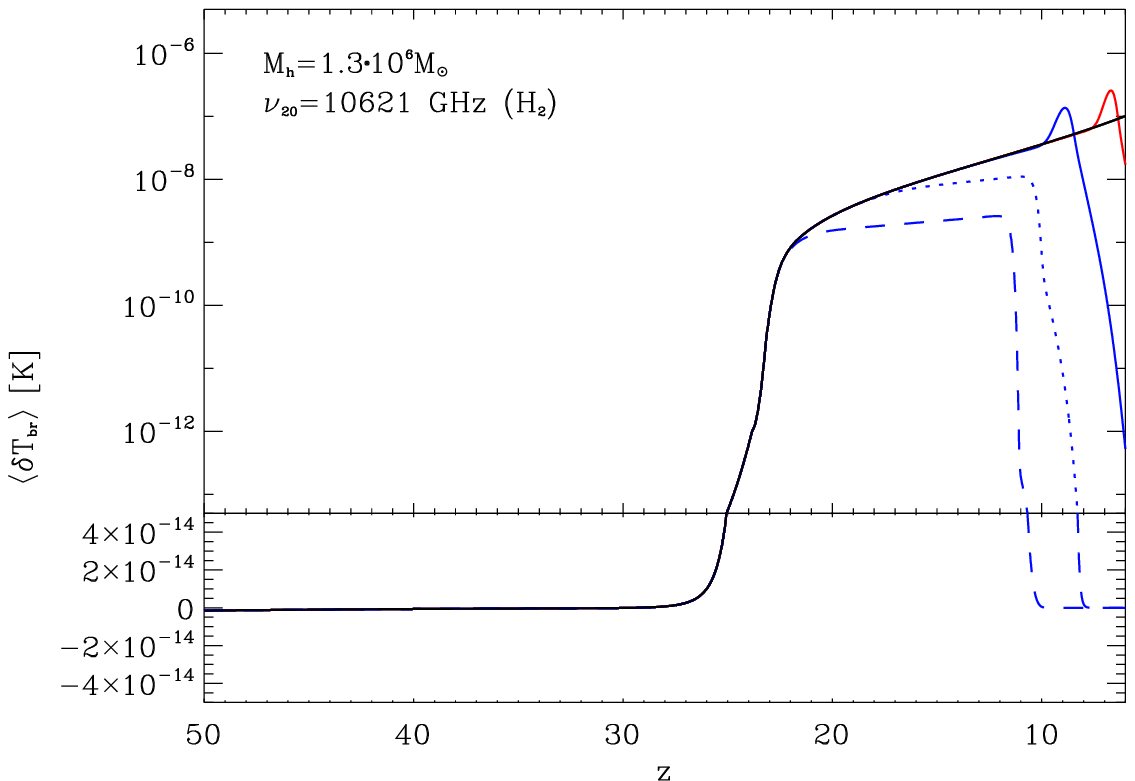}
\includegraphics[width=0.49\textwidth]{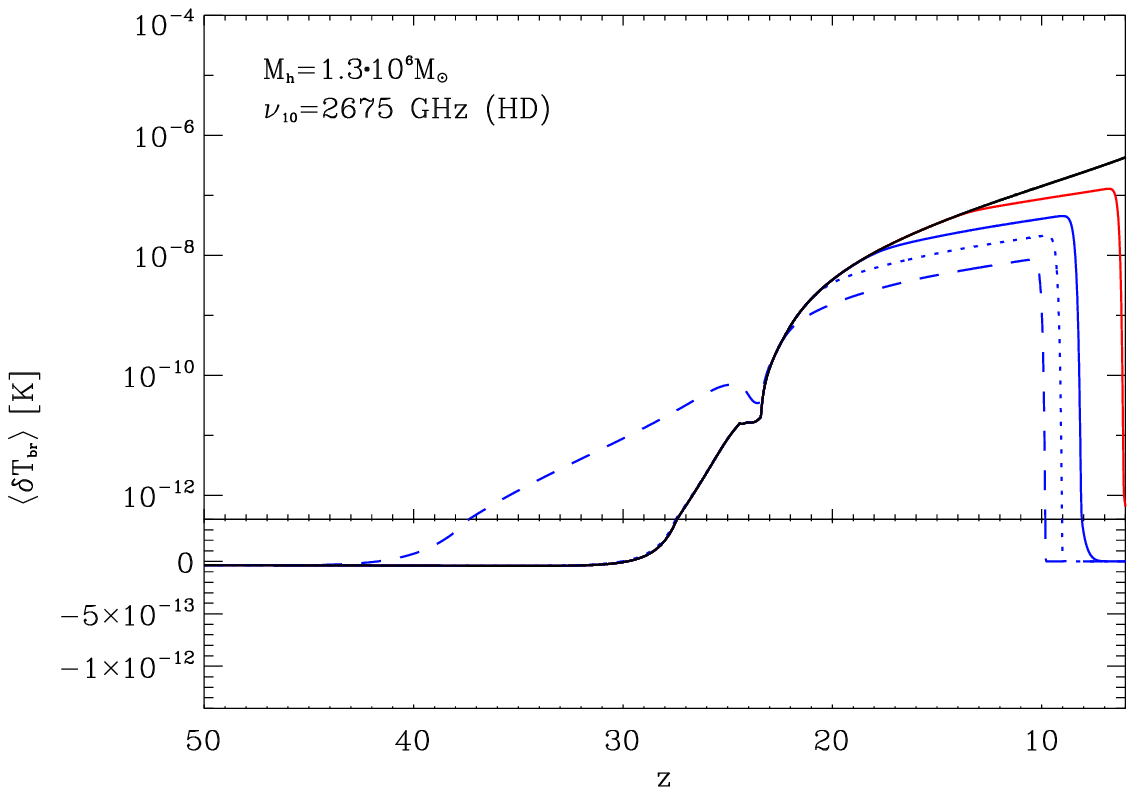}
\includegraphics[width=0.49\textwidth]{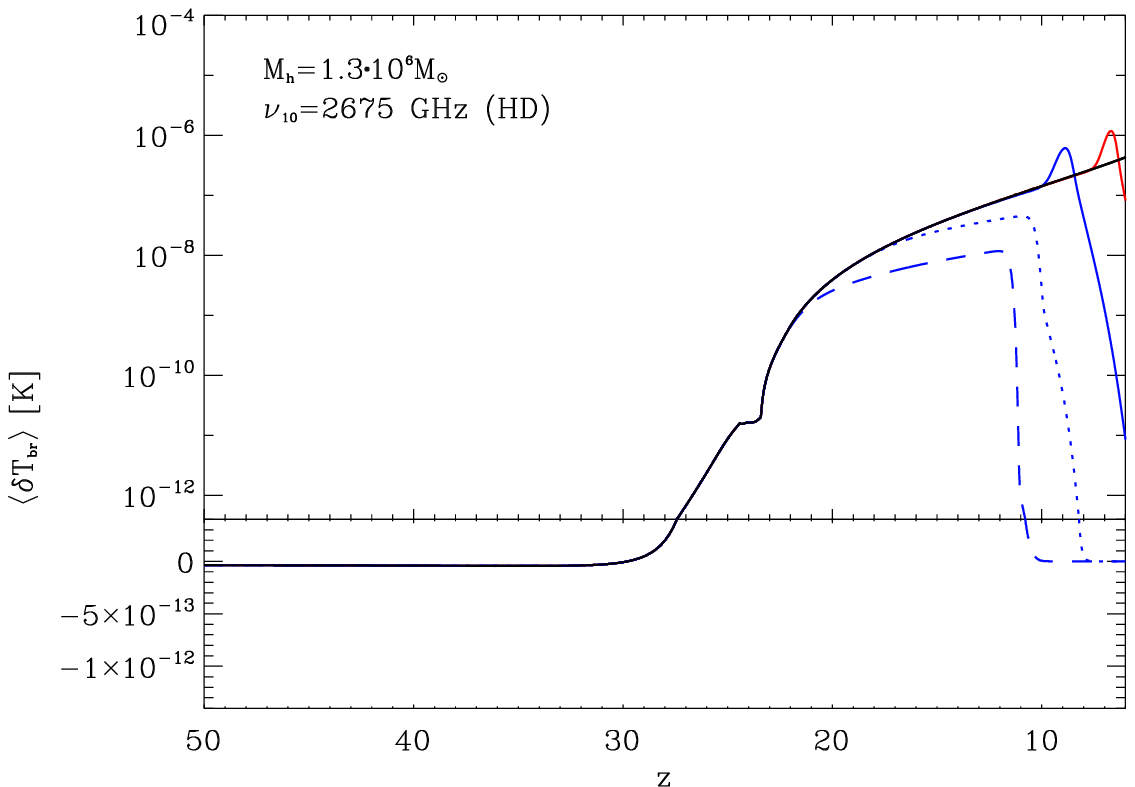}
\includegraphics[width=0.49\textwidth]{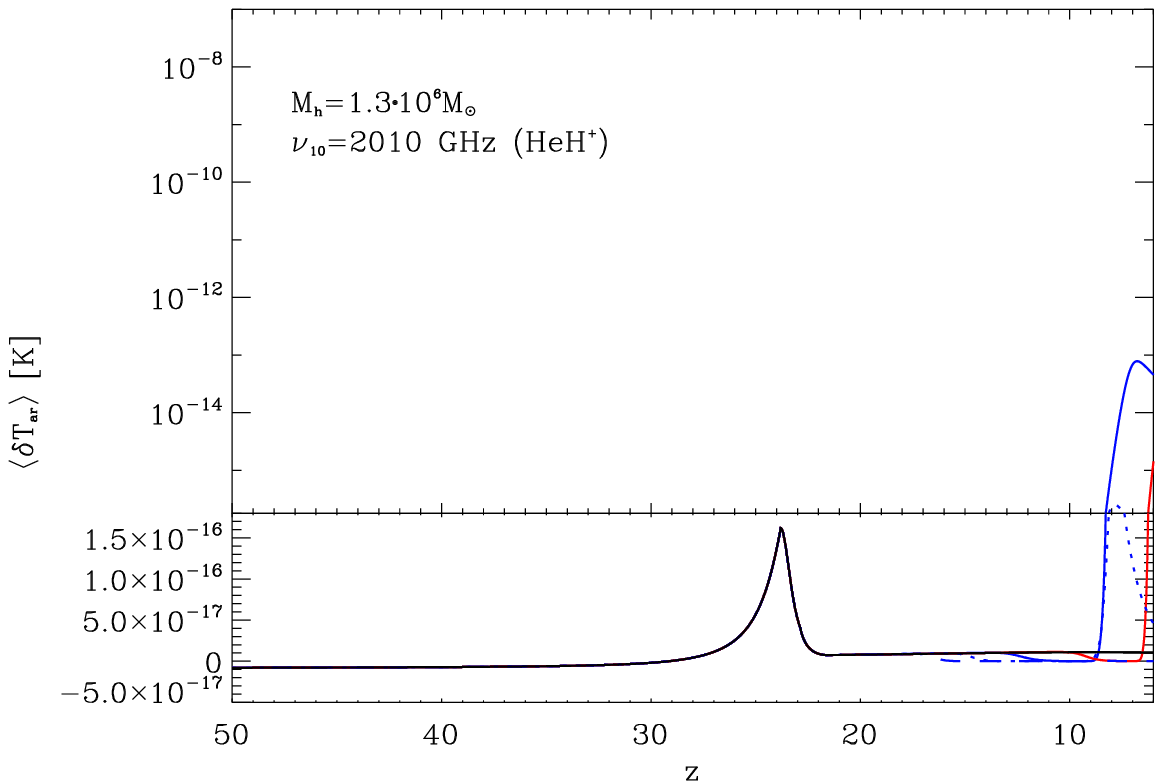}
\includegraphics[width=0.49\textwidth]{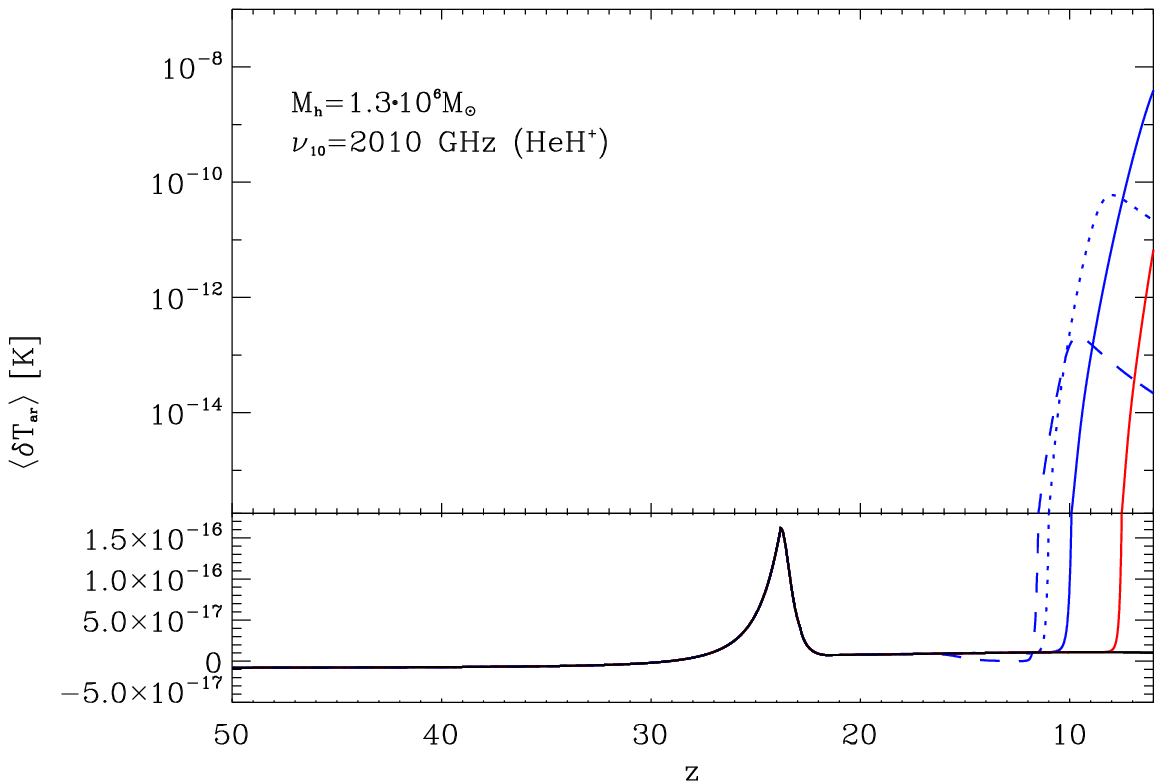}
\caption{Differential brightness temperature of hot halos ($T_b\approx2280$ K) in the base rotational lines H$_2$, HD and HeH$^+$ (from top to down) with and without of the first light: solid dark line is without the first light; solid red and blue lines with background are the first light that reionize the medium at z=6 and z=8, accordingly; dotted and dashed blue lines with the first light with an energy density that is $10^3$ and $10^6$ times higher than the background one ($z_{rei}=8$). In the left column, the first light is Ia, in the right one - Ib.}
\label{Tbr-hh}
\end{figure*}

Solution of the system of equations for populations of rotational levels with $J=0-5$ gives us the possibility to compute the excitation temperature from the $i$-level to the $j$-level:
\begin{equation}
T_{ex}=\frac{h_P\nu_{ij}}{k_B}/\ln{\frac{n_j}{n_i}\frac{g_i}{g_j}},\nonumber 
\end{equation}
where $h_P$ is the Planck constant, $k_B$ is the Boltzmann constant, and $g_i$, $g_j$ are statistical weights of levels.
This, in turn, makes it possible to calculate the optical depths in the rotational lines, $\nu_{ij}$, along the diameter of the spherical homogeneous top-hat halo,
\begin{equation}
\tau_{ij}=1.55\cdot10^{47}n_j\frac{g_i}{g_j}\frac{A_{ij}}{\nu_{ij}^3}\sqrt{\frac{m_A}{T_k}}\left[1-\exp{\left(-\frac{h_P\nu_{ij}}{k_B T_{ex}}\right)}\right]r_h,
\label{tau}
\end{equation}
where $m_A$ is the atomic number of a molecule, $r_h$ is the radius of the halo in units of kpc, and the rest values are in the CGS system. The results show that the maximum values of $\tau^{H_2}_{20}$, $\tau^{H_2}_{31}$, $\tau^{HD}_{10}$ and $\tau^{HeH^+}_{10}$ after virialization are in the range of $10^{-7}\div10^{-9}$, which means that such halos are optically thin. Our estimation shows that they are also optically thin  in the frequencies of photo-dissociations of molecules ($\tau_{pd}\le10^{-6}$). Thus, the use of the optically thin halo approximation in the molecular lines and in the continuum is justified. Since the column densities of the molecules $N(H_2),\,N(HD),\,N(HeH^+)\ll10^{20}$ cm$^2$, the self-shielding effect is not important for such halos \citep{Draine1996}.

Traditionally, the intensity of microwave radiation is expressed in units of brightness temperature:
\begin{equation}
T^{br}_{ij}\equiv\frac{I_{\nu_{ij}}}{2k_B\nu_{ij}^2}. \nonumber 
\end{equation}
Since the luminescence of the Dark Ages and Cosmic Dawn halos in the lines of the first molecules occurs on the cosmic microwave background at the same frequencies, it is reasonable to calculate the differential brightness temperature:
\begin{equation}
\delta T^{br}_{ij}\equiv T^{br}_{ij}-T_{CMB}.\nonumber
\end{equation}
In the case of an optically thin top-hat halo the radiative transfer equation gives the next expression for the differential brightness temperature; for details, see \cite{Novosyadlyj2020}:
\begin{equation}
\delta T^{br}_{ij}\approx\frac{5\cdot10^{36}n_iA_{ij}}{(1+z)\nu_{ij}^2}\sqrt{\frac{m_A}{T_b}}\left[\frac{\exp{\left(\frac{h_P\nu_{ij}}{k_BT_{CMB}}\right)}-\exp{\left(\frac{h_P\nu_{ij}}{k_BT_{ex}}\right)}}{\exp{\left(\frac{h_P\nu_{ij}}{k_BT_{CMB}}\right)} -1}\right]r_h.
\label{dTb2} 
\end{equation}
Using it we compute the evolution of luminescence of warm and hot halos in the rotational lines of molecules H$_2$, HD, and HeH$^+$ during the Dark Ages and Cosmic Dawn with the different energy densities of the first light. The results are presented in  Figs. \ref{Tbr-wh} and \ref{Tbr-hh} for the same models of the first light as in Fig.~\ref{h_Ia-Ib}. 

We can see that the luminescence of a single halo in the rotational lines of H$_2$ and HD molecules appears after the turn-around point of seed perturbation, when adiabatic heating of baryon matter becomes noticeable and disappears before hydrogen reionization of the intra-halo medium caused by photodissociation of molecules. It has formed a plateau with sharp uphill and downhill. Their detection could be very informative about the first halos, first molecules, and first light. Unfortunately, the amplitude of the differential brightness temperature is too low to be detected by current telescopes. However, in the case of the ``hotter'' first light with lower energy density, the differential brightness temperature of hot halos can reach a few hundredths or even tenths of a microkelvin (Fig.~\ref{Tbr-hh}, right panels in the top and middle rows), at the  sensitivity level of existing and planned advanced telescopes for the microwave range.

The time dependence of the luminescence of a single halo in the rotational lines of HeH$^+$ molecular ions is as interesting with regard to both warm halos as well as hot ones. It has three phases of sharp changes: increasing after the turn-around point of seed perturbation, decreasing during the Cosmic Dawn caused by photodissociation of molecules in the case of the warm halo (Fig.~\ref{Tbr-wh}, bottom row), or by collision in the case of the hot halo and a noticeable increase during the reionization. Thus, the halo could exhibit luminescence at the reionization epoch in the rotational lines of HeH$^+$ molecular ions only (from the studied ones, of course). Again, however, the amplitude of differential brightness temperature in the basic rotational lines is too low to be detected by existing telescopes. 

The magnitude of fluxes in the rotational lines of the first molecules obtained in this work and in \cite{Novosyadlyj2020} without reionizing by the first light practically coincides with that predicted in the early works  \citep{Dubrovich1977,Maoli1994,Maoli1996,Kamaya2002,Ripamonti2002,Kamaya2003,Omukai2003,Mizusawa2004,Mizusawa2005,Dubrovich2008} despite completely different source models. The main conclusion about their marginal detectability by modern and planned telescopes coincides with the conclusions of these and other authors. The novel result presented here is the luminescence of H$_2$ and HD molecules in the lines long before the reionization of atomic hydrogen and their disappearance due to photodissociation of these molecules shortly prior. The halo luminescence with $M\propto10^{10}-10^{13}$ M$_\odot$ at $7\le z\le 24$ has been estimated using GIZMO cosmological simulations (see, e.g., \cite{Liu2019}). The interpolation of their results to smaller halo masses shows a good agreement (up to orders of magnitude of fluxes) with our results in the models without the first light. The fact that  there is no disappearance of H$_2$ emission at $z\le8-9$ due to photodissociation can be explained by the underestimation of UV radiation, since the model of reionization is not  self-consistent in this case.

\medskip

The results presented here were obtained within the framework of the simplest models of halo and first light, which can be far from real, as well as a limited set of chemical reactions and energy levels of molecules. Complicating the models by expanding the number of reactions and energy transitions in the molecules will be a task for further research. 

\section{Discussion and conclusions} 

We analyzed the ionization of the hydrogen atoms and the formation and destruction of the first molecules in the inter-proto-galactic medium of Cosmic Dawn, assuming the thermal energy distribution of the background radiation of the first sources of light. The spectral energy distribution at those epochs was presented by the Planck function with a certain temperature dependent on redshift, along with a dilution factor constrained by data about the redshift of reionization. The latter was assumed to be  $x_{HII}=0.5$ at $6\le z\le8$. We used the simplest  model of the first light with a minimal number of parameters (temperature and dilution coefficient). Such radiation is not in thermodynamic equilibrium but features Planck energy distribution, which essentially simplifies the computations of all activation and deactivation and photodissociation of atoms and molecules, according to the generalized formula (5). It corresponds to the following scenario: the sources of the first light, proto-, and PopIII stars, evolve, while the average energy density and temperature of their radiation increase according to Equations (\ref{Tfl}) and (4). The parameters in Equation (\ref{Tfl}) could be constrained using the observational data on the luminescence of halos in the lines of molecules, which we have evaluated here. However, in this paper, we set them arbitrarily in order to clarify the effect of the star formation rate on our estimates. Mathematically, the problem has consisted of the numerical integration of the system of kinetic equations of the ionization and recombination of atoms and molecules, as well as the formation and destruction of molecules and their ions together with the equations of expansion of the Universe and the thermal energy balance for the baryonic component. We only took into account the most important reactions (Table \ref{tab1}) that are connected with the most abundant atoms and molecules: atomic hydrogen H, helium He, deuterium D,  molecular hydrogen H$_2$, hydrogen deuteride molecule HD, and helium hydride ion HeH$^+$. The obtained results lead to conclusions that provide a clearer understanding of the possible physical conditions at  Cosmic Dawn and the epoch of reionization. Our conclusions are summarized as follows: 

1. The number densities of hydrogen atoms, kinetic temperature, and the radiation field in the inter-proto-galaxy medium of Cosmic Dawn are such that the atoms are in the base state mainly and the case of A of photoionization is realized (Fig.~1). 

2. The thermal radiation with a temperature of $T_{fl}=5000$ K and dilution factor of $\alpha_{fl}\sim 2\div5\cdot10^{-10}$ ionizes the atomic hydrogen up to $x_{HII}=0.5$ at $6\le z\le8$. At the current epoch, the spectral energy density of such radiation is a few orders lower than the spectral energy density of CIB at frequencies of <30 THz ($\lambda>10\,\mu$m). At higher frequencies, it would be detected by the current radio telescopes. For higher radiation temperatures, the dilution factor is lower and the current energy density of such radiation becomes too low for its detection. For the first light with comoving radiation temperature, $T_{fl}=10000$ K, which ionizes the inter-proto-galaxy medium at $6\le z\le8,$ the dilution factor is $\sim 10^{-17}$. Its spectral flux is in the range of the frequency maximum at the current epoch, which is $\sim$ Jy, and masks in the Wien range of CIB. The range of values for the dilution factor $\alpha_{fl}$ is narrow for any fixed radiation temperature of the first light, $T_{fl}$, at the moment of reionization, but the dependence  $\alpha_{fl}(T_{fl})$ is steep since the potential of ionization of the atomic hydrogen is in the Wien range of spectral energy distribution (Figs.~\ref{flI}-\ref{xHII}).

3.The first light heats the baryonic gas in the inter-proto-galaxy medium up to dozens or hundreds of kelvins at $z_{rei}$, which depends on its radiation temperature and dilution factor. In any case, the change of baryonic temperature $T_b$ becomes noticeable not long before the ionized hydrogen fraction reaches 1/2. The different history of increasing of radiation temperature, $T_{fl}$, forms the temperature history of the baryonic component at that time (Fig.~\ref{roT}).

4. The number densities of negative hydrogen ion H$^-$ and molecular ions H$_2^+$ start to decrease long before reionization, since the thresholds energy of photodissociation for them are essentially lower than the potential of ionization of the atomic hydrogen (Figs.~\ref{nIab}-\ref{nIcd}).

5. The most abundant  molecules in the Dark Ages, namely, H$_2$ and HD, are destroyed over the course of the Cosmic Dawn not long before hydrogen reionization since the threshold of photodissociation for them is lower than the potential of ionization of atomic hydrogen (Figs.~\ref{nIab}-\ref{iIa-d}).

6. The dependence of the number density of the helium hydride ion  HeH$^+$ on redshift in the Cosmic Dawn is intriguing via the competition of photodissociation by the first light and its formation in the reaction $\mathrm{He + H^+ \rightarrow HeH^+ + \gamma}$ when the number density of H$^+$  increase. The number density of HeH$^+$ decreases in the models of the first light with a large number density of photons, and increases for diluted first light with higher radiation temperature (Figs.~\ref{nIab}-\ref{iIa-d}.) We consider whether these molecules can survive in the remote corners of the intergalactic medium? This is an interesting task for further research with more realistic models of the spectral energy distribution of radiation in the different periods of the evolution of the Universe. 

7. The time dependence of the luminescence of a single top-hat halo in the rotational lines of H$_2$ and HD molecules has a form of plateau with sharp uphill and downhill and maximal amplitude of a differential brightness temperature of $\sim10^{-10}-10^{-8}$ K for warm halos and $\sim10^{-8}-10^{-6}$ K for hot ones (Figs. \ref{Tbr-wh}--\ref{Tbr-hh}). The beginning of the plateau is connected with the formation and adiabatic heating of the forming halo, while the ending of the plateau is caused by photodissociation of the molecules before reionization.

8. The time dependence of the luminescence of a single halo in the rotational lines of HeH$^+$ molecular ions has three phases of sharp changes: increasing during virialization halo (heating), decreasing during Cosmic Dawn (photodissociation or collision destruction of molecules), and increasing during the reionization epoch. The maximal amplitude of differential brightness temperature does not exceed $10^{-7}$ K. 

9. The detection of the luminescence of the Dark Ages and Cosmic Dawn halos in the rotational lines of the first molecules could provide very important information about formation of the first halos, first molecules, and the first sources of light.

\begin{acknowledgements}
This work was performed within the project of Ministry of Education and Science of Ukraine ``Modeling the luminosity of elements of the large-scale structure of the early universe and the remnants of galactic supernovae and the observation of variable stars'' (state registration number 0122U001834) and was supported also by the International Center of Future Science of Jilin University. We would also like to thank the Armed Forces of Ukraine for their resilience and courage which saved us and  provided security to finalize this work during the unprovoked invasion russian army into Ukraine.
\end{acknowledgements}

\begin{appendix}
\section{Rates of photorecombination and effective cross-sections of photoionization for HI and DI }
Before the epoch of reionization, during epochs of the Dark Ages and Cosmic Dawn, hydrogen atoms were mainly in the ground electron state. To obtain the total rate of electron recombination on the second level, we use the analytical approximation of the total rate of recombination on all levels by \cite{Verner1996}:
\begin{equation}
\alpha_r(T)=\frac{a}{\sqrt{T/T_1}\left(1+\sqrt{T/T_1}\right)^{1-b}\left(1+\sqrt{T/T_2}\right)^{1+b}}\nonumber,
\end{equation}
with the coefficients $a=7.982\cdot10^{-11}$, $b=0.748$, $T_1=3.148$, $T_2=7.036\cdot10^{5}$, and the approximation of the rate of recombination on the base level by \cite{Ferland1992}:
\begin{eqnarray}
\alpha(1,T)&=&10^{F(1,T)}/T,\nonumber\\
F(1,T)&=&\frac{a_1+c_1x+e_1x^2+g_1x^3+i_1x^4}{1+b_1x+d_1x^2+f_1x^3+h_1x^4},\quad x=\lg{T,}\nonumber
\end{eqnarray}
with refined values of approximation coefficients from the publicly available code Cloudy \citep{Ferland2017} $a_1=-10.781359$, $b_1=-0.38790005$, $c_1=4.6748385$, $d_1=0.063898347$, $e_1=-0.87297469$, $f_1=-5.0923340\cdot10^{-3}$,  $g_1=0.081445619$,  $h_1=2.4935490\cdot10^{-4}$, and $i_1=-3.9043790\cdot10^{-3}$. So, the expected value of the recombination rate coefficient for case B is as follows:
\begin{equation}
\alpha_{rB}(T)=\alpha_r(T)-\alpha(1,T).\nonumber
\end{equation}
The testing shows that this approximation of the case B of recombination coefficient $\alpha_{rB}(T)$ matches the values $\sum_{n=2}^{400}\alpha(n,T)$ in the range of electron temperature $3\le T\le10^5$K computed by Cloudy with an accuracy of about $0.01\%$.

\begin{figure}[ht!]
\includegraphics[width=0.48\textwidth]{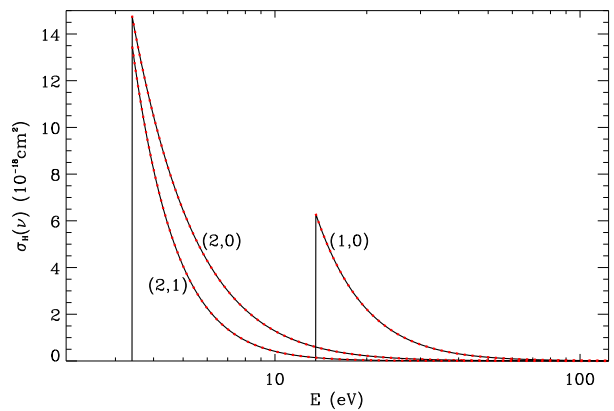} 
\caption{Effective cross-sections of photoionization of hydrogen atoms H from the base level ($n=1,\,l=0$) and second one ($n=2,\,l=0$), and ($n=2,\,l=1$). Solid lines show the numerical computations, dotted ones the analytical approximations.}
\label{csH}
\end{figure}

The rate of photoionization of atomic hydrogen from the level $n$ by radiation with intensity, $I(\nu),$ we can compute, for the known effective cross-section of photoionization from this level  $\sigma_n(\nu),$ according to \cite{Abel1997}, as follows:
\begin{equation}
R_{nc}=4\pi\int_{\nu_{nc}}^\infty\sigma_n(\nu)\frac{I(\nu)}{h\nu}d\nu,\hskip4.9cm (A1)\nonumber
%\label{rate}
\end{equation}
where $\nu_{nc}$ is the threshold frequency of photoionization from level $n$. For the intergalactic cold rarefied medium of the Dark Ages and Cosmic Dawn, photoionization goes mainly from base ($n=1,\, l=0$) or second metastable levels ($n=2,\,l=0$). We use the analytical approximation proposed by \cite{Verner1996b} with a small correction: 
\begin{equation}
 \sigma_n(E)=\sigma_0(x-1)^2x^{P/2-5.5}/\left(1+\sqrt{x/y_a}\right)^P/\left(1+\alpha(E-E_{min})\right), \hskip0.4cm (A2)\nonumber
%\label{cs}
\end{equation}
where $x=\frac{E}{E_0}$, $\sigma_0$, $E=h\nu$, $E_0$, $P$, $y_a$ and $\alpha$ are the best-fit parameters, which we re-determined using the open source codes by \cite{Bauman2001} for computations the cross-sections of photoionization of hydrogen from the base level ($n=1,\,l=0$) and the second one ($n=2,\,l=0$) and ($n=2,\,l=1$) shown in Fig.~\ref{csH} as solid lines. The best-fit coefficients and accuracy of approximation are presented in Table \ref{coef}.
\begin{table*}[ht!]
\begin{center}
\caption{Coefficients of approximation (A2) for cross-sections of photoionization of HI from the first and second energy levels.}
\begin{tabular} {cccccccccc}
\hline
\hline
   \noalign{\smallskip}
$n$&$l$&$E_{min}$&$E_{max}$&$E_0$&$\sigma_0$&$y_a$&$P$&$\alpha$&Max. dev.\\
 \noalign{\smallskip} 
& &eV &eV &eV&$10^{-18}\,cm^2$&&&&\% \\
 \noalign{\smallskip} 
\hline
   \noalign{\smallskip} 
 1&0&13.6&$10^6$&1.0235&$1.1083\cdot10^{-14}$&23.424&2.3745&0&0.015\\
 2&0&3.4&$10^6$&0.065935&$1.6555\cdot10^{-14}$&85.921&4.8780 &0&0.015\\
 2&1&3.4&10&0.65253&$6.1775\cdot10^{-12}$&$6.4642\cdot10^{-06}$&0.95820&0&0.64\\
  & &10 &40&5.5590&$3.1454\cdot10^{-12}$&$6.1461\cdot10^{-11}$&1.0346&2.4402&1.4\\
  & &40&200&1.0363&$1.1128\cdot10^{-11}$&$4.5855\cdot10^{-10}$&0.85601&0.77849&11\\ 
  &&200&690& 0.011796&$1.6424\cdot10^{-9}$&$5.1814\cdot10^{-24}$&-0.022506&1.0960&10\\ 
    \noalign{\smallskip}    
  \hline
  \hline
\end{tabular}
\label{coef}
\end{center}
\end{table*}

The effective cross-sections of photoionization of HI from the base and second levels are shown in Fig.~\ref{csH} by the dotted red line.
We assume that the rates of photorecombination and effective cross-sections of photoionization for deuterium atoms are the same as for atomic hydrogen.

\section{Rates of photodissociation of molecules H$_2$ and HD }
\begin{figure*}[ht!]
\includegraphics[width=0.49\textwidth]{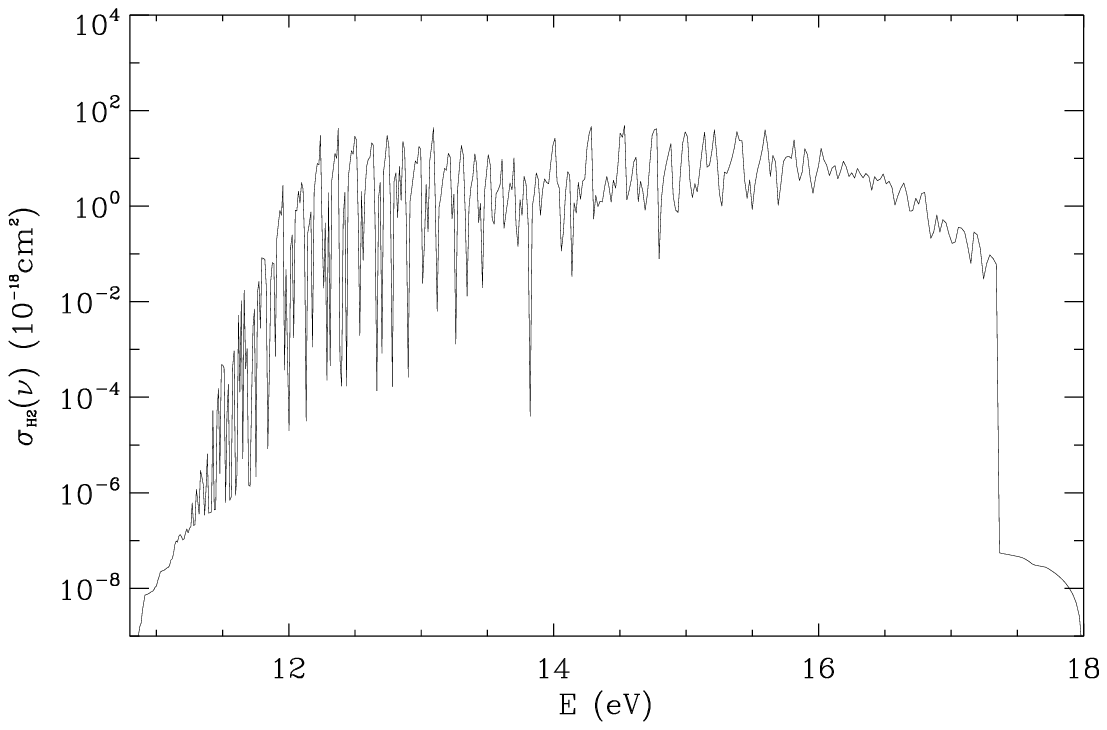}
\includegraphics[width=0.47\textwidth]{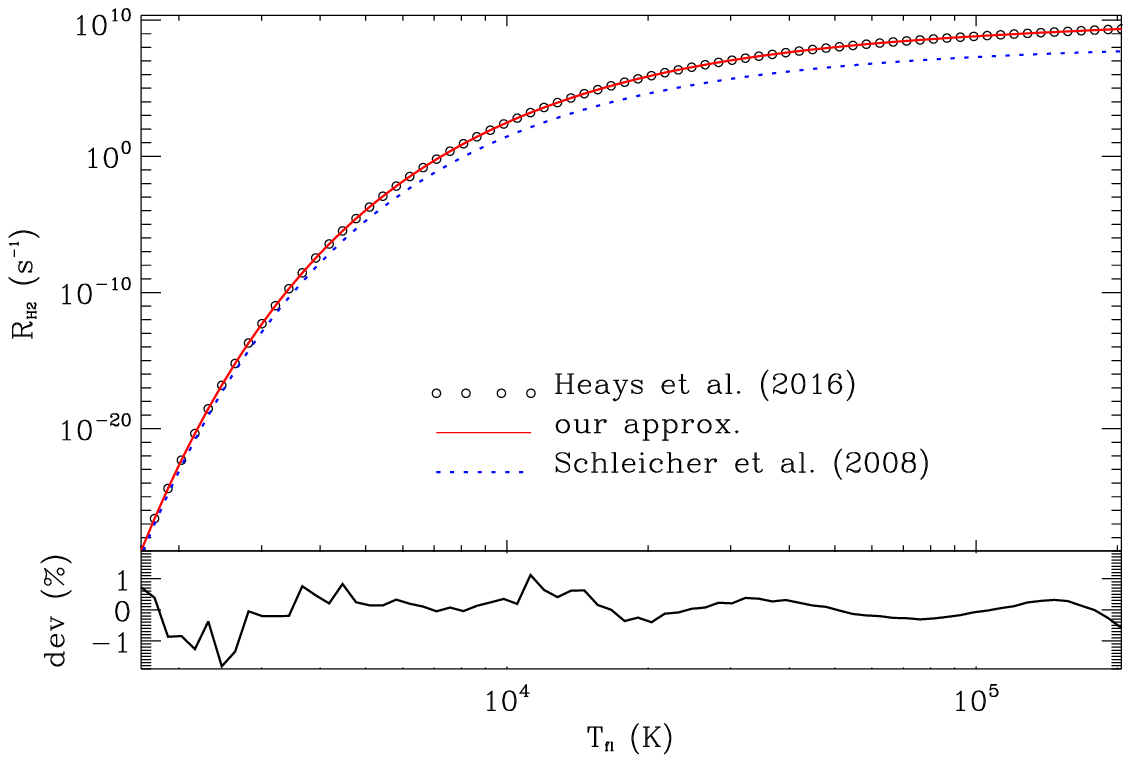}
\caption{Effective cross-sections of hydrogen molecule H$_2$ photodissociation from \cite{Heays2017} shown in the left panel and the rate of dissociation computed here in the right panel. The solid red line is the analytical approximation (B1), the dotted blue line is from \cite{Glover2007,Schleicher2008}. The deviation of analytical approximation (B1) from the numerical results is shown in the low panel.}
\label{RdH2}
\end{figure*}

At the Cosmic Dawn, energetic photons appear and they are capable to ionize of atoms and molecules and dissociate the last. The threshold of direct photodissociation of H$_2$ (H$_2$ + $\gamma$ $\rightarrow$ H + H) is 4.53 eV, so, it can be important long before the reionization. The physics and computations of the cross-section of this reaction are well described in \cite{Glover2007,Heays2017} (see also citations therein). Here, we use the photodissociation cross-section of molecule H$_2$ computed by \cite{Heays2017} and presented on the site of Leiden University\footnote{https://home.strw.leidenuniv.nl/$~$ewine/photo/cross\_sections.html} in the table form with high wavelength resolution ($4\cdot10^{-5}\mu$m). We illustrate these data in Figure \ref{RdH2} using low wavelength resolution table data ($0.1\mu$m).
We computed the rate of dissociation according to (A1) with $I(\nu)=B_{\nu}(T)$ for range of temperatures $10^3<T<2\cdot10^5$ K. The result is shown in the right panel of Fig.~\ref{RdH2} by open circles. We also propose the analytical approximation of this dependence by the formula: 
\begin{equation}
R_{H_2}(T)=aT^b(1+cT^d)/\exp(f/T),\hskip4.0cm (B1)\nonumber
%\label{RdH2a}
\end{equation}
with the best-fit parameters $(a,\,b,\,c,\,d,\,f)=(2394.9,\,0.62681,\,2.4635,\,0.56957,\,140050)$ for $T\le15000$ K and $(a,\,b,\,c,\,d,\,f)=(6.4471\cdot10^7,\,0.322,\,6.5341\cdot10^{-8},\,1.4752,\,153570)$ for $T>15000$ K. Its deviation from the numerical result is not larger than 1\% in the interested temperature range (bottom small panel of the left figure). We present also in the main panel of the right figure the analytical approximation by \cite{Glover2007,Schleicher2008} for comparison. We can see that in the low temperature range ($T<5000$ K), they are close. We assume here that the rate of direct photodissociation of molecule HD (HD + $\gamma$ $\rightarrow$ D +
\end{appendix} H) is the same as for molecule H$_2$.

\begin{thebibliography}{}
\bibitem[Abel et al.(1997)]{Abel1997} Abel, T., Anninos, P., Zhang, Yu., Norman, M.L. 1997, New Astron., 2, 181 
\bibitem[Banados et al.(2018)]{Banados2018} Banados, E., Venemans, B. P., Mazzucchelli, C., et al. 2018, Nature, 553, 473
\bibitem[Barkana et al.(2001)]{Barkana2001} Barkana, R., Loeb, A. 2001, Phys. Rep., 349, 125
\bibitem[Bates \& Opik(1968)]{Bates1968} Bates, D.R. \& Opik, U. 1968, J. Phys. B, 1, 543
\bibitem[Bauman et al.(2001)]{Bauman2001} Bauman, R.P., Ferland, G.J., MacAdam, K.B. 2001, Bul. AAS, 33, 1331
\bibitem[de Bernardis et al.(1993)]{deBernardis1993} de Bernardis, P., Dubrovich, V., Encrenaz, P. et al. 1993, A\&A, 269, 1 
\bibitem[Bowman et al. (2018)]{Bowman2018} Bowman, J.D., Rogers, A.E.E., Monsalve, R.A. et al. 2018, Nature, 555, 67 
\bibitem[Bouwens et al.(2015)]{Bouwens2015} Bouwens, R. J., Illingworth, G. D., Oesch, P. A., et al. 2015, ApJ, 811, 140
\bibitem[Bromm \& Yoshida(2011)]{Bromm2011} Bromm, V. \& Yoshida, N. 2011, Ann. Rev. A\&A, 49, 373
\bibitem[Cooke et al.(2018)]{Cooke2018} Cooke, R.J., Pettini, M., \& Steidel, C.C. 2018, ApJ, 855, 102
\bibitem[Davies et al.(2018)]{Davies2018} Davies, F. B., Hennawi, J. F., Bañados, E., et al. 2018, ApJ, 864, 142
\bibitem[Draine \& Bertoldi(1996)]{Draine1996} Draine, B.T. \& Bertoldi, F. 1996, ApJ, 468, 269 
\bibitem[Dubrovich(1977)]{Dubrovich1977} Dubrovich, V. K. 1977, Sov. Astron. Lett., 3, 128
\bibitem[Dubrovich et al.(2008)]{Dubrovich2008}  Dubrovich, V., Bajkova, A., Khaikin, V. B. 2008,  New Astronomy, 13, 28
\bibitem[Ferland et al.(1992)]{Ferland1992} Ferland, G.J., Peterson, B.M., Horne, K. et al. 1992, ApJ, 387, 95
\bibitem[Ferland et al.(2017)]{Ferland2017} Ferland, G.J., Chatzikos, M., Guzmán, F. et al. 2017, Rev. Mexic. Astron. Astrof., 53, 385
\bibitem[de Jong(1972)]{deJong1972} de Jong, T., 1972, A\&A, 20, 263
\bibitem[Galli \& Palla(1998)]{Galli1998} Galli, D., Palla, F. 1998, A\&A, 335, 403
\bibitem[Glover \& Jappsen(2007)]{Glover2007} Glover, S.C. \& Jappsen, A.-K. 2007, ApJ, 666, 1
\bibitem[Gosachinskij et al.(2002)]{Gosachinskij2002} Gosachinskij, I. V., Dubrovich, V. K., Zhelenkov, S. R., Il’in, G. N., \& Prozorov, V. A. 2002, Astron. Rep., 46, 543
\bibitem[Grachev \& Dubrovich(1991)]{Grachev1991} Grachev, S.I., Dubrovich, V.K., 1991, Astrophysics, 34, 124
\bibitem[Heays et al.(2017)]{Heays2017} Heays, A.N., Bosman, A.D. \& van Dishoeck, E.F. 2017, A\&A 602, A102
\bibitem[Hills et al. (2018)]{Hills2018} Hills, R., Kulkarni, G., Meerburg, P.D., Puchwein, E. 2018, Nature, 564, E32
\bibitem[Holliday et al.(1971)]{Holliday1971} Holliday, M.G., Muckerman, J.T., Friedman, L. 1971, J. Chem. Phys., 54, 1058
\bibitem[Kamaya \& Silk(2002)]{Kamaya2002} Kamaya, H. \& Silk, J. 2002, MNRAS,  332, 251
\bibitem[Kamaya \& Silk(2003)]{Kamaya2003} Kamaya, H. \& Silk, J. 2003, MNRAS, 339, 1256
\bibitem[Karpas et al.(1979)]{Karpas1979} Karpas, Z., Anicich, V., Huntress, W.T. 1979, J. Chem. Phys., 70, 2877
\bibitem[Kogut et al.(2019)]{Kogut2019} Kogut, A., Abitbol, M.H., Chluba,   J. 2019, arXiv:1907.13195
\bibitem[Kulinich et al.(2020)]{Kulinich2020} Kulinich, Yu., Novosyadlyj, B., Shulga, V., Han, W. 2020, Phys. Rev. D, 101, 083519
\bibitem[Launey et al.(1991)]{Launey1991} Launay, J.M., Le Dourneuf, M., Zeippen, C.J. 1991, A\&A 252, 842
\bibitem[Liu et al.(2019)]{Liu2019} Liu, B., Jaacks, J., Finkelstein, S.L. and Bromm, V. 2019, MNRAS 486, 3617
\bibitem[Maio et al.(2022)]{Maio2022} Maio, U., Péroux, C. and Ciardi, B. 2022, A\&A, 657, A47
\bibitem[Maoli et al.(1994)]{Maoli1994} Maoli, R., Melchiorri, F., \& Tosti, D. 1994, ApJ, 425, 372
\bibitem[Maoli et al.(1996)]{Maoli1996} Maoli, R., Ferrucci, V., Melchiorri, F., \& Tosti, D. 1996, ApJ, 457, 1
\bibitem[Mason et al.(2018)]{Mason2018} Mason, C. A., Treu, T., Dijkstra, M., et al. 2018, ApJ, 856, 2
\bibitem[Matsuda et al.(1971)]{Matsuda1971} Matsuda, T., Sato, H. and Takeda, H. 1971, Prog. Theor. Phys., 46, 416
\bibitem[Mizusawa et al.(2004)]{Mizusawa2004} Mizusawa, H.,  Nishi, R. \& Omukai, K. 2004, Publ. Astron. Soc. Japan, 56, 487
\bibitem[Mizusawa et al.(2005)]{Mizusawa2005} Mizusawa, H., Omukai, K. \& Nishi, R. 2005, Publ. Astron. Soc. Japan, 57, 951
\bibitem[Moseley et al.(1970)]{Moseley1970} Moseley, J., Aberth, W., Peterson, J.A. 1970, Phys. Rev. Lett., 24, 435
\bibitem[Novosyadlyj et al.(2016)]{Novosyadlyj2016} Novosyadlyj, B., Tsizh, M., Kulinich, Yu. 2016, Gen. Relativ. Grav., 48, 30
\bibitem[Novosyadlyj et al.(2017)]{Novosyadlyj2017} Novosyadlyj, B., Sergijenko, O., Shulga, V.M. 2017, Kinematics and Physics of Celestial Bodies, 33, 255
\bibitem[Novosyadlyj et al.(2018)]{Novosyadlyj2018} Novosyadlyj, B., Shulga, V., Han, W., Kulinich, Yu., \& Tsizh, M. 2018, ApJ, 865, 38
\bibitem[Novosyadlyj et al.(2020)]{Novosyadlyj2020} Novosyadlyj, B., Shulga, V., Kulinich, Yu., Han, W. 2020, ApJ, 888, 27
\bibitem[Omukai \& Kitayama(2003)]{Omukai2003} Omukai, K. \& Kitayama, T. 2003, ApJ, 599, 738
\bibitem[ONeil \& Reinhardt(1978)]{ONeil1978} ONeil, S.V., Reinhardt, W.P. 1978, J. Chem. Phys., 69, 2126
\bibitem[Osterbrock(1989)]{Osterbrock1989} Osterbrock, D.E., 1989, Astrophysics of Gaseous Nebulae and Active Galactic Nuclei, University Science Books, 408 p.
\bibitem[Peebles(1968)]{Peebles1968} Peebles P.J.E. 1968, ApJ, 153, 1
\bibitem[Peimbert et al.(2016)]{Peimbert2016} Peimbert, A., Peimbert, M., \& Luridiana, V. 2016, Rev. Mex. Astron. Astrofis.,52, 419
\bibitem[Persson et al.(2010)]{Persson2010} Persson, C. M., R. Maoli, R., Encrenaz, P. et al. 2010, A\&A, 515, A72
\bibitem[Planck Collaboration(2020a)]{Planck2020a} Planck Collaboration, 2020, A\&A,  641, A1
\bibitem[Planck Collaboration(2020b)]{Planck2020} Planck Collaboration, 2020, A\&A,  641, A6 
\bibitem[Poulaert et al.(1978)]{Poulaert1978} Poulaert, G., Brouillard, F., Claeys, W., McGowan J.W., van Wassenhove G. 1978, J. Phys. B, 11, L671 
\bibitem[Puy et al.(1993)]{Puy1993} Puy, D., Alecian, G., Le Bourlot, J. et al. 1993, A\&A, 267, 337
\bibitem[Ripamonti et al.(2002)]{Ripamonti2002} Ripamonti, E., Haardt, F., Ferrara, A. and Colpi, M. 2002, MNRAS, 334, 401
\bibitem[Roberge \& Dalgarno(1982)]{Roberge1982} Roberge, W. \& Dalgarno, A., 1982, ApJ, 255, 489
\bibitem[Schleicher et al.(2008)]{Schleicher2008} Schleicher, D.R.G., Galli, D., Palla, F. et al. 2008, A\&A, 490, 521
\bibitem[Schneider et al.(1994)]{Schneider1994} Schneider, I.F., Dulieu, O., Giusti-Suzor, A., Roueff, E. 1994, ApJ, 424, 983 
\bibitem[Seager et al.(1999)]{seager1999} Seager, S., Sasselov,  D. D., Scott, D. 1999, ApJ, 523, L1 
\bibitem[Seager et al.(2000)]{seager2000} Seager, S., Sasselov,  D. D., Scott,  D. 2000, ApJ, Suppl. Ser., 128, 407
\bibitem[Smith et al.(1982)]{Smith1982} Smith, D., Adams, N.G., Alge, E. 1982, ApJ, 263, 123
\bibitem[Stancil(1994)]{Stancil1994} Stancil, P.C. 1994, ApJ, 430, 360
\bibitem[Stancil et al.(1993)]{Stancil1993} Stancil, P.C., Babb, J.F., Dalgarno, A., 1993, ApJ, 414, 672
\bibitem[Verner \& Ferland(1996)]{Verner1996} Verner, D.A. \& Ferland, G.J. 1996, ApJSS, 103, 467
\bibitem[Verner et al.(1996)]{Verner1996b} Verner, D.A., Ferland, G.J., Korista, K.T., Yakovlev, D.G. 1996, ApJ, 465, 487
\bibitem[Vonlanthen et al.(2009)]{Vonlanthen2009} Vonlanthen, P., Rauscher, T., Winteler, C., Puy, D., Signore, M., Dubrovich, V. 2009, A\&A, 503, 47
\bibitem[Watson et al.(1978)]{Watson1978} Watson, W.D., Christensen, R.B., Deissler, R.J. 1978, A\&A, 69, 159
\bibitem[Yousif \& Mitchell(1989)]{Yousif1989} Yousif, F.B. \& Mitchell, J.B.A. 1989, Phys. Rev. A, 40, 4318
\bibitem[Zygelman\& Dalgarno(1990)] {Zygelman1990} Zygelman, B., Dalgarno, A. 1990, ApJ, 365, 239 
\end{thebibliography}
\end{document}